\documentclass[a4paper,journal]{IEEEtran}
\usepackage{sty/26XX_NFFB}

\def\BibTeX{{\rm B\kern-.05em{\sc i\kern-.025em b}\kern-.08em
    T\kern-.1667em\lower.7ex\hbox{E}\kern-.125emX}}
 
\begin{document}

\title{On Near-Far-Field Boundaries in Wireless Systems}

\author{Alexander Stutz-Tirri, \IEEEmembership{Member, IEEE}, Ahmad Dkhan, \IEEEmembership{Member, IEEE}, \\Hadi Sarieddeen, \IEEEmembership{Senior Member, IEEE}, and Christoph Studer, \IEEEmembership{Senior Member, IEEE}
\thanks{First submitted on April 29, 2026. The work of A. Stutz-Tirri and C. Studer was funded in part by armasuisse. 
The work of A. Dkhan and H. Sarieddeen was funded in part by the American University of Beirut's University Research Board and Vertically Integrated Projects Program. 
The work of C. Studer was also funded by the Swiss State Secretariat for Education, Research, and Innovation (SERI) under the SwissChips initiative, the Swiss National Science Foundation (SNSF) grant 200021\_207314, by a CHIST-ERA grant for the project CHASER (CHIST-ERA-22-WAI-01) through the SNSF grant 20CH21\_218704, and by the European Commission within the context of the project 6G-REFERENCE (6G Hardware Enablers for Cell Free Coherent Communications and Sensing), funded under EU Horizon Europe Grant Agreement 101139155.}
\thanks{A. Stutz-Tirri is with the Department of Information Technology and Electrical Engineering, ETH Zurich, Switzerland (e-mail: alstutz@ethz.ch).}
\thanks{A. Dkhan is with the Department of Electrical and Computer Engineering, American University of Beirut, Lebanon (e-mail: amd53@mail.aub.edu).}
\thanks{H. Sarieddeen is with the Department of Electrical and Computer Engineering, American University of Beirut, Lebanon (e-mail: hadi.sarieddeen@aub.edu.lb).}
\thanks{C. Studer is with the Department of Information Technology and Electrical Engineering, ETH Zurich, Switzerland (e-mail: cstuder@ethz.ch).}
\thanks{The authors thank Dr.~Raphael Rolny and Dr.~Gian Marti for helpful discussions and suggestions on this paper. }}

\maketitle


\begin{abstract}

Near-field (NF) multi-antenna wireless communication and sensing have attracted growing research interest in recent years. A core question in this area is how to determine whether a wireless system is operating in the NF or far-field (FF) region. In this work, we propose a framework grounded in Maxwell’s equations to analyze the transition region between the NF and FF, following the IEEE definition that specifies where the NF ends and the FF begins. Using this framework, we (i) compare a variety of traditional and recently introduced single-letter distance thresholds---often referred to as near–far-field boundaries---and (ii) conduct numerical experiments with both single- and multi-antenna wireless systems and with analytical models as well as full-wave electromagnetic simulations. Our results indicate that all of the considered single-letter distance thresholds are insufficient to predict the transition region between the NF and FF regions. Moreover, we highlight several important caveats associated with frequently (and recently) used NF and FF concepts. 
\end{abstract}



\section{Introduction}

\begin{figure}[tp]
        
    {
    \small

    \tdplotsetmaincoords{80}{108}
    
    \begin{tikzpicture}[yscale=1,xscale=-1]

        \clip (-7cm,-4.2cm) rectangle (2cm, 2cm);  
    
        \begin{scope}[shift={(6mm,0)}]
            \fill[black!1] (-23.3,-4) circle (20);
            \fill[black!2] (-23.35,-4) circle (20);
            \fill[black!3] (-23.4,-4) circle (20);
            \fill[black!4] (-23.45,-4) circle (20);
            \fill[black!5] (-23.5,-4) circle (20);
            \fill[black!6] (-23.55,-4) circle (20);
            \fill[black!7] (-23.6,-4) circle (20);
            \fill[black!8] (-23.65,-4) circle (20);
            \fill[black!9] (-23.7,-4) circle (20);
            \fill[black!10] (-23.75,-4) circle (20);
        \end{scope}
    
        \begin{scope}[shift={(-.17,-.15)}]
            \draw [black, line width=.3mm] plot [smooth,samples=200, tension=0.6] coordinates { 
           (-.5, .55) (0, .5) (.5, .8) (1, .7) (1.2, .2) (1.5, .1) (1.6, -.2) (1.3, -.5) (.4, -.35) (-.3, -.5) (-.7, -.4) (-.76, -.2) (-.8, -.1) (-.9, 0) (-1.15, 0.1) (-1.2, 0.45) (-1, .6) (-.5, .55)};
           \draw [black, line width=.3mm] plot [smooth,samples=200, tension=1] coordinates { 
           (-.5, .5585) (-.6, 0) (-.5, -.48) }; 
           \draw [black, line width=.3mm] plot [smooth,samples=200, tension=1] coordinates { 
           (-.95, .6) (-1.03, .3) (-1, 0.03) }; 
           \draw [black, line width=.3mm] plot [smooth,samples=200, tension=1] coordinates { 
           (0, .5) (-.08, 0) (0, -.43) }; 
           \draw [black, line width=.3mm] plot [smooth,samples=200, tension=1] coordinates { 
           (.6, .81) (.4, .2) (.5, -.36) }; 
           \draw [black, line width=.3mm] plot [smooth,samples=200, tension=1] coordinates { 
           (1.05, .63) (.92, 0) (1, -.465) };   
           \draw [black, line width=.3mm] plot [smooth,samples=200, tension=1] coordinates { 
           (1.48, .12) (1.46, -.2) (1.5, -.4) };  
           \draw [black, line width=.3mm] plot [smooth,samples=200, tension=1] coordinates { 
           (-.9, 0) (.3, -.1) (1.57, 0) }; 
           \draw [black, line width=.3mm] plot [smooth,samples=200, tension=1] coordinates { 
           (-1.17, .5) (0, .37) (1.1, .5) }; 
    
        \end{scope}
    
        \draw[white, line width=.8mm] (0, 0) -- (-5*.5, -3*.5);
        \draw[line width=.3mm] (0, 0) -- (-5*.8*.7, -3*.8*.7);
    
        \begin{scope}[shift={(5*0.8*0.425,3*0.8*0.425)}]
            \draw[line width=.3mm] (-5*.92, -3*.92) -- (-5*.95, -3*.95);
            \draw[line width=.3mm] (-5*.97, -3*.97) -- (-5*1, -3*1);
    
            \draw[line width=.3mm] (-5*1.02, -3*1.02) -- (-5*1.4, -3*1.4);
    
            \draw[line width=.3mm] (-5*1.42, -3*1.42) -- (-5*1.45, -3*1.45);
            \draw[line width=.3mm] (-5*1.47, -3*1.47) -- (-5*1.5, -3*1.5);
        \end{scope}

        \draw[color=black!10, line width=.8mm] (-5*.87,-3*.87) circle (.3); 
        \draw[color=white, line width=.8mm, shift={(5/4,3/4)}] (-5*.6,-3*.6) circle (.3);
    
        \draw[white, line width=.8mm] (-2.01,-1.2) -- (-0.97,-3.12);
        \draw[white, line width=.8mm] (-1.71,-.752) -- (.44,-.969);
    
        \draw[black!10, line width=.8mm] (-3.81,0) -- (-4.05,-2.65);
        \draw[black!10, line width=.8mm] (-6.18,-0.9) -- (-4.59,-2.79);
        
        \draw[black!80, line width=.3mm] (-3.81,0) -- (-4.05,-2.65);
        \draw[black!80, line width=.3mm] (-6.18,-0.9) -- (-4.59,-2.79);
        
        \draw[black!80, line width=.3mm] (-2.01,-1.2) -- (-0.97,-3.12);
        \draw[black!80, line width=.3mm] (-1.71,-.753) -- (.44,-.969);
        
        \draw[color=black!80, line width=.3mm, shift={(5/4,3/4)}] (-5*.6,-3*.6) circle (.3); 
        \draw[color=black!80, line width=.3mm] (-5*.87,-3*.87) circle (.3); 
        \draw[line width=.8mm, black!10, shift={(-5*.87,-3*.87)}] (117:.3) arc (117:175:.3);
        
        \begin{scope}[shift={(1.2,-1.75)}]
            \draw[color=black!80, fill=white, line width=.3mm] (-5*.2,-3*.2) circle (1.4); 
            \draw[line width=.3mm*1.5] (-5*.1, -3*.1) -- (-5*.3, -3*.3);
            \draw[line width=.3mm*1.5] (-5*.32, -3*.32) -- (-5*.35, -3*.35);
            \draw[line width=.3mm*1.5] (-5*.08, -3*.08) -- (-5*.05, -3*.05);
    
            \begin{scope}[shift={((-5*.2, -3*.2))}]
    
                \begin{scope}
                    \draw[color=white, line width=.8mm] (168:1.4) arc (168:181:1.4);
                    \draw[color=white, line width=.8mm] (132:1.4) arc (132:151:1.4);
                    \draw[color=white, line width=.8mm] (46:1.4) arc (46:72:1.4);
                \end{scope}
    
            
                \begin{scope}[tdplot_main_coords]
                    \small
                    \draw[eth_blue, line width=.25mm, arrows={-Triangle[angle=35:1.5mm]}] (0,0,0) -- (0,-1,.6) node[shift={(-3mm,2.6mm)}] {$Z_0^{-\frac{1}{2}}\phv{E}_{\up{F}\up{F}}$};
                    \draw[eth_blue, line width=.25mm, arrows={-Triangle[angle=35:1.5mm]}] (0,0,0) -- (0,.6,1) node[shift={(-0.7mm,2.7mm)}] {$Z_0^{\frac{1}{2}}\phv{H}_{\up{F}\up{F}}$};
                    \draw[eth_green, line width=.25mm, arrows={-Triangle[angle=35:1.5mm]}] (0,0,0) -- (0,-1,.4) node[shift={(1mm,-2.7mm)}] {$Z_0^{-\frac{1}{2}}\phv{E}$};
                    \draw[eth_green, line width=.25mm, arrows={-Triangle[angle=35:1.5mm]}] (0,0,0) -- (0,.4,.0) node[shift={(-3.5mm,0.4mm)}] {$Z_0^{\frac{1}{2}}\phv{H}$};
                \end{scope}
    
                \node[] at (.2mm,-2.8mm) {$\vect{r}$};
                \fill (0,0) circle[radius=.8mm];
            \end{scope}

        \end{scope}
    
        \begin{scope}[shift={(-4.2,.7)}]
            \draw[color=black!80, fill=black!10, line width=.3mm] (-5*.2,-3*.2) circle (1.4); 
    
            \begin{scope}[shift={(-5*.2,-3*.2)}]
                \draw[color=black!10, line width=.8mm] (147:1.4) arc (147:181:1.4) ;
            \end{scope}

            \begin{scope}[shift={(-5*.2,-3*.2)}]
                \draw[color=black!10, line width=.8mm] (48:1.4) arc (48:80:1.4) ;
            \end{scope}

            \draw[line width=.3mm*1.5] (-5*.1, -3*.1) -- (-5*.3, -3*.3);
            \draw[line width=.3mm*1.5] (-5*.32, -3*.32) -- (-5*.35, -3*.35);
            \draw[line width=.3mm*1.5] (-5*.08, -3*.08) -- (-5*.05, -3*.05);
        
            \begin{scope}[shift={((-5*.2, -3*.2))}]
                
                \begin{scope}[tdplot_main_coords]
                    \small

                    \draw[eth_blue, line width=.25mm, arrows={-Triangle[angle=35:1.5mm]},shift={(0.15mm,0.19mm)}] (0,0,0) -- (0,-.5,.3) node[shift={(6.35mm,-2mm)}] {$Z_0^{-1/2}\phv{E}_{\up{F}\up{F}}$};
                    \draw[eth_blue, line width=.25mm, arrows={-Triangle[angle=35:1.5mm]},shift={(-0.2mm,0.15mm)}] (0,0,0) -- (0,.3,0.5) node[shift={(-1.5mm,3mm)}] {$Z_0^{1/2}\phv{H}_{\up{F}\up{F}}$};
                
                    \draw[eth_green, line width=.25mm, arrows={-Triangle[angle=35:1.5mm]}] (0,0,0) -- (0,-.5,.3) node[shift={(5mm,2mm)}] {$Z_0^{-1/2}\phv{E}$};
                    \draw[eth_green, line width=.25mm, arrows={-Triangle[angle=35:1.5mm]}] (0,0,0) -- (0,.3,0.5) node[shift={(-2.6mm,7.5mm)}] {$Z_0^{1/2}\phv{H}$};
                \end{scope}
                
                \fill (0,0) circle[radius=.8mm];
                \node[] at (2mm,-2.5mm) {$\vect{r}_\up{FF}$};
            
            \end{scope}
        \end{scope}

        \fill[] (-5*.87,-3*.87) circle[radius=.4mm] node[shift={(1.4mm,1.5mm)}] {$\vect{r}_\up{FF}$};

        \fill[shift={(5/4,3/4)}] (-5*.6, -3*.6) circle[radius=.4mm] node[shift={(-.6mm,-1.5mm)}] {$\vect{r}$};

        \begin{scope}[shift={(-3*0.2,5*0.2)}]
            \draw [white, line width=.8mm] (-5*-0.0,-3*-0.0) -- (-5*0.35,-3*0.35);
            \draw [black, line width=.3pt] (-5*-0.03,-3*-0.03) -- (-5*0.38,-3*0.38);
            \draw [line width=.3pt, arrows = {-Stealth[inset=0, length=6pt, angle'=25]}] (-5*-0,-3*-0) -- (-5*0.35,-3*0.35);
            \draw [line width=.3pt, arrows = {-Stealth[inset=0, length=6pt, angle'=25]}] (-5*0.35,-3*0.35) -- (-5*-0,-3*-0);
            \node[shift={(1mm,2mm)}] at (-5*0.175,-3*0.175) {\small $r$};
        \end{scope}

        \draw [white, line width=.8mm] (0,0) -- (-3*0.1,5*0.1);
        \draw [black, line width=.15mm] (0,0) -- (-3*0.23,5*0.23);

        \begin{scope}[shift={(-5*0.35,-3*0.35)}]
            \draw [white, line width=.8mm] (-3*0.04,5*0.04) -- (-3*0.1,5*0.1);
            \draw [black, line width=.15mm] (0,0) -- (-3*0.23,5*0.23);
        \end{scope}

        \pgftext{\includegraphics[width=.28cm]{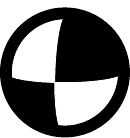}} at (0pt,0pt);  
        
        \node[] at (.1,1.4) {\small reference point};
    
        \node[] at (0.8,1.75) {\small antenna system};
        \draw [black, line width=.15mm] plot [smooth,samples=200, tension=.8] coordinates { 
        (1.5, 1.6) (1.2, .9) (.95,.6)};
    
        \node[] at (-2.93,-3.3) {\small test line in direction ($\theta,\varphi$)};
    
        \draw [white, line width=.8mm] plot [smooth,samples=200, tension=.8] coordinates { 
           (.2, 1.2) (.05, .5) (.05, .16)};
        \draw [black, line width=.15mm] plot [smooth,samples=200, tension=.8] coordinates { 
           (.2, 1.2) (.05, .5) (.05, .16)};
    
        \draw [black, line width=.15mm] plot [smooth,samples=200, tension=1] coordinates { 
        (-3.3, -3.1) (-3, -2.5) (-2.7, -1.7)};

        \node[fill=black!10] at (-5,-4.15+.2) {\color{black!60}\small far-field region};

    \end{tikzpicture}
    \caption{Basic concept of our framework: Illustration of the relationship between the antenna system, the reference point, the test line, the real electromagnetic field characterized by $\phv{E}$ and $\phv{H}$, and the auxiliary field functions, $\phv{E}_\up{FF}$ and $\phv{H}_\up{FF}$. 
It should be noted that, unlike in the region close to the antenna system, the real fields coincide with the auxiliary fields in the far-field region. 
The prefactors related to the free-space impedance $Z_0$ are introduced to weight deviations of the electric and magnetic fields equally.}

\vspace{0.2cm}

    \label{fig:concept_framework}

    \begin{tikzpicture}[yscale=.7,xscale=1.1]

        \clip (-1,-.8) rectangle (7cm, 3.3cm); 
        \begin{scope}[shift={(-1.47,0)}]
            \fill[black!.1] (4.7,-.35) rectangle (10,2.4);
            \fill[black!2] (4.75,-.35) rectangle (10,2.4);
            \fill[black!3] (4.8,-.35) rectangle (10,2.4);
            \fill[black!4] (4.85,-.35) rectangle (10,2.4);
            \fill[black!5] (4.9,-.35) rectangle (10,2.4);
            \fill[black!6] (4.95,-.35) rectangle (10,2.4);
            \fill[black!7] (5,-.35) rectangle (10,2.4);
            \fill[black!8] (5.05,-.35) rectangle (10,2.4);
            \fill[black!9] (5.1,-.35) rectangle (10,2.4);
            \fill[black!10] (5.15,-.35) rectangle (10,2.4);
        \end{scope}
    
        \draw[black!20, line width=.2mm] (0,.5) -- (6,.5);
        \draw[black!20, line width=.2mm] (0,1) -- (6,1);
        \draw[black!20, line width=.2mm] (0,1.5) -- (6,1.5);
        \draw[black!20, line width=.2mm] (0,2) -- (6,2);
    
        \draw[black!20, line width=.2mm] (1,0) -- (1,2);
        \draw[black!20, line width=.2mm] (2,0) -- (2,2);
        \draw[black!20, line width=.2mm] (3,0) -- (3,2);
        \draw[black!20, line width=.2mm] (4,0) -- (4,2);
        \draw[black!20, line width=.2mm] (5,0) -- (5,2);
        \draw[black!20, line width=.2mm] (6,0) -- (6,2);

        \draw[line width=.3mm, arrows={-Triangle[angle=35:1.5mm]}] (0,0) -- (0,2.3) node[shift={(10mm,3mm)}] {\color{eth_black} approximation error $\epsilon_{\theta,\phi}(r)$};
        \draw[line width=.3mm, arrows={-Triangle[angle=35:1.5mm]}] (0,0) -- (6.35,0) node[shift={(2mm,0mm)}] {\color{eth_black} $r$};
    
        \draw [line width=.3mm] plot [smooth,samples=200, tension=.6] coordinates { 
           (0, 2)  (0.3, 2) (1.2, 1.9) (1.6, .6) (2, 0.8) (2.7, 0.16) (3.2, 0.04) (3.7, 0.0012) (4.2, 0.001) (6.3, 0.00)};
    
        \draw [line width=.3mm, eth_purple] (3.5,-0.35) -- (3.5,2.2);
        \draw[->, line width=.2mm] (0,-.4) -- (0,-1mm);
        \node[] at (0,-.55) {\small reference point};
        \node[] at (3.5,-.55) {\color{eth_purple}\small boundary under test};
    
    \end{tikzpicture}\caption{Sketch of how a result of the proposed framework could look: 
    The approximation error $\epsilon$, which quantifies the deviation of the real electromagnetic field components $\phv{E}$ and $\phv{H}$ from the auxiliary field functions $\phv{E}_\up{FF}$ and $\phv{H}_\up{FF}$, takes on a significant positive value near the reference point before eventually converging to zero. 
    A ``good'' near-far-field boundary under test should enable accurate prediction of the position at which this convergence occurs. 
    In the example shown, the boundary under test appears to perform well.}
    \label{fig:concept_example_plot}
    }
\end{figure}

Wireless communication and sensing systems typically operate in their far-field (FF) region, i.e., they interact with objects located at distances sufficiently large so that the fields emitted by the system exhibit an angular distribution that is essentially independent of distance.\footnote{To be specific, in the FF region, the electromagnetic fields that the wireless system emits into free space can be accurately approximated as in~\fref{eq:far_field_E} and~\fref{eq:far_field_H}.} 
However, with the emergence of very large antenna arrays that enable fine-grained beamforming capabilities, this paradigm is changing. 
See, for instance, the contributions in the recent \emph{Special Issue on Near-Field Signal Processing} of the \emph{IEEE Signal Processing Magazine}, particularly the Editors’ introductory article~\cite{elbir_special_issue_on_near_field_signal_processing}.
As the near-field (NF) region of an antenna system expands with increasing array size, wireless systems are also beginning to operate within their NF region. 
Consequently, signal processing techniques specialized for NF communication and sensing are becoming increasingly important.
Two core questions arise in this research area: (i) Where does the NF–FF transition region lie, and (ii) how can one determine its location for a given antenna system?
The first question is answered by the \emph{IEEE Standard for Definitions of Terms for Antennas}~\cite[Sec.~4]{IEEE_standard_for_definitions_of_terms_for_antennas_2013}, as we will detail in~\fref{sec:field_regions_intro}. 
Nonetheless, several recent papers attempt to answer the question of where the NF–FF transition lies by perpetuating the misconception that the NF and FF regions of an antenna system can be distinguished solely based on whether the electromagnetic field exhibits a plane- or spherical-wave character---we will provide a concrete counterexample to this misconception in~\fref{sec:field_regions_intro}.

To answer the second question, several single-letter thresholds, so-called near-far-field boundaries, have been proposed in the literature~\cite{lu_zeng_communicating_with_extremely_large_array_surface,zhang_zhang_liu_di_codebook_design_and_beam_alignment_for_ios_aided_communications,renwang_shu_meixia_applicable_regions_of_spherical_and_plane_wave_modes_for_extremely_large_scale_array_communications,daei_fodor_skoglund_when_near_becomes_far, lu_dai_near_field_channel_estimation_in_mixed_los_nlos_environments_for_extremely_large_scale_mimo_systems, sun_how_to_differentiate_between_near_far_field}.
However, little is known about how well such thresholds locate the NF-FF transition region of general antenna systems; this is particularly the case because such thresholds are often \emph{not} derived from first principles, i.e., from Maxwell's equations. 

\subsection{Contributions}
In this paper, we address the second core question concerning the location of the NF–FF transition region by proposing a framework for analyzing this transition region for arbitrary antenna systems. 
Our framework, illustrated in~\fref{fig:concept_framework}, is (i) grounded in Maxwell’s equations, (ii) builds on the \emph{IEEE} definitions~\cite[Sec.~4]{IEEE_standard_for_definitions_of_terms_for_antennas_2013} of the NF and FF regions, and (iii) capable of dealing with calculated, simulated, or measured electromagnetic fields. 
We leverage our framework to evaluate a selection of traditional and recently introduced single-letter near-far-field boundaries for various antenna systems, including arrays of infinitesimal dipoles, half-wave dipoles, and patch antennas. 
Based on our results, we discuss the accuracy and usefulness of such existing near-far-field boundaries. 
Finally, we highlight several important and recently recurring misconceptions associated with antenna field regions.

\subsection{Notation}
We use boldface for vectors (e.g., $\vect{a}$); 
we use sans-serif (e.g.,~$\phs{s}$) and sans-serif boldface (e.g.,~$\phv{s}$) to represent phasors\footnote{The phasor $\phs{s}=\frac{1}{\sqrt{2}} A e^{j\phi_0}$ represents the sinusoidal time function $s(t)=A \cos(2\pi f t+\phi_0)$, $t\in\mathbb{R}$, which oscillates at frequency $f\in\mathbb{R}_{>0}$ with a phase of~$\phi\in\mathbb{R}$.}  and vectors containing phasors, respectively. 
The superscripts~$^\T$ and~$^\He$ represent the transpose (e.g., $\vect{a}^\T$) and conjugate transpose (e.g., $\vect{a}^\He$), respectively. 
The Euclidean norm is denoted with~\mbox{$\|\cdot\|_2$}. 
The cross product of two three-dimensional vectors $\vect{a}$ and $\vect{b}$ is denoted by $\vect{a}\times\vect{b}$. 
We refer to the $n$th element of a vector $\vect{a}$ as $\vect{a}_n$. 
Given $N\in\mathbb{N}$, we define the set $[N]\triangleq\{1, \ldots,N\}$. 
We define the imaginary unit by $j^2=-1$. 
To simplify notation, we refrain from explicitly including physical units, except when specifying numerical values. All physical equations are formulated in SI units.

\section{The Field Regions of Antennas}\label{sec:field_regions_intro}

The space surrounding an antenna (or an antenna system) is typically divided into the NF region and the FF region~\cite[Sec.~2.2.4]{balanis_antenna_theory_edition_3}.
This separation is (i) based on the properties of the electromagnetic field that the antenna system emits in free space and (ii) seamless between the two regions, i.e., without any sharp boundary. 
According to the \emph{IEEE Standard for Definitions of Terms for Antennas}~\cite[Sec.~4]{IEEE_standard_for_definitions_of_terms_for_antennas_2013}, 
the \emph{far-field region} is ``that region of the field of an antenna where the angular field distribution is essentially independent of the distance from a specified point in the antenna's region;'' 
and the \emph{near-field region} is ``that part of space between the antenna and the far-field region.'' 
Throughout this paper, we adhere to these definitions. 
In recent years, various publications have proposed near–far-field boundaries that depend on the properties of both a receiving and a transmitting antenna system.\footnote{For example, some papers estimate the near–far-field boundary based on the ``aperture'' of the transmitting \emph{and} receiving antenna arrays.} 
In doing so, however, the field regions (of an antenna system) are treated as dependent on the surroundings\footnote{The surroundings include, for example, other antenna systems.} of this antenna system, which is in disagreement with the IEEE definitions cited above. 
\begin{rem}\label{rem:intrinsic_property}
    The field regions of an antenna (system) are \emph{intrinsic properties} of that antenna (system). Specifically, the NF and FF regions are independent of the larger system in which the antenna (system) is embedded.
\end{rem}
\begin{rem}\label{rem:multiple_tranceivers}
    It follows directly from~\fref{rem:intrinsic_property} that, in a communication system comprising two transceivers, each transceiver's antenna system has its own NF and FF regions. 
    Consequently, two near–far-field boundaries exist. 
    Moreover, system A can be located in system B’s far-field region while system B lies in system A’s near-field region, and vice versa. 
\end{rem}

\begin{figure}[tp]

    \def\centerarc[#1](#2)(#3:#4:#5)
    { \draw[#1] ($(#2)+({#5*cos(#3)},{#5*sin(#3)})$) arc (#3:#4:#5); }

    \begin{tikzpicture}
        
        \begin{scope}[scale=1]
    
            \clip (-.5cm,-2cm) rectangle (8.5cm, 1.5cm);
    
            \begin{scope}[shift={(48,4)}]
                \fill[black!1] (-24.6,-4) circle (20);
                \fill[black!2] (-24.5,-4) circle (20);
                \fill[black!3] (-24.4,-4) circle (20);
                \fill[black!4] (-24.3,-4) circle (20);
                \fill[black!5] (-24.2,-4) circle (20);
                \fill[black!6] (-24.1,-4) circle (20);
                \fill[black!7] (-24.0,-4) circle (20);
                \fill[black!8] (-23.9,-4) circle (20);
                \fill[black!9] (-23.8,-4) circle (20);
                \fill[black!10] (-23.7,-4) circle (20);
            \end{scope}
    
            \centerarc[line width=.2mm, black!80](-60.5,0)(-.68:.68:64)
            \centerarc[line width=.2mm, black!80](-28.2,0)(-1.47:1.47:32)
            \centerarc[line width=.2mm, black!80](-11.9,0)(-3.15:3.15:16)
            \centerarc[line width=.2mm, black!80](-3.6,0)(-6.7:6.7:8)
            \centerarc[line width=.2mm, black!80](-1.3,0)(-9.5:9.5:6)
            \centerarc[line width=.2mm, black!80](0,0)(-12:12:5)
    
            \centerarc[line width=.2mm, black!80](1,0)(-15:15:4.3)
            \centerarc[line width=.2mm, black!80](1,0)(-15:15:4.3)
            \centerarc[line width=.2mm, black!80](1,0)(-15:15:4.6)
            \centerarc[line width=.2mm, black!80](1,0)(-15:15:4.9)
            \centerarc[line width=.2mm, black!80](1,0)(-15:15:5.2)

            \centerarc[line width=.2mm, black!80](1,0)(-15:15:5.5)
            \centerarc[line width=.2mm, black!80](1,0)(-15:15:5.8)
            \centerarc[line width=.2mm, black!80](1,0)(-15:15:6.1)
            \centerarc[line width=.2mm, black!80](1,0)(-15:15:6.4)
            \centerarc[line width=.2mm, black!80](1,0)(-15:15:6.7)
            \centerarc[line width=.2mm, black!80](1,0)(-15:15:7.0)
            \centerarc[line width=.2mm, black!80](1,0)(-15:15:7.3)

            \draw[line width=.2mm, black!80] (1.7,-.7) -- (1.7,.7);
            \draw[line width=.2mm, black!80] (2,-.7) -- (2,.7);
            \draw[line width=.2mm, black!80] (2.3,-.7) -- (2.3,.7);
            \draw[line width=.2mm, black!80] (2.6,-.7) -- (2.6,.7);
            \draw[line width=.2mm, black!80] (2.9,-.7) -- (2.9,.7);
            \draw[line width=.2mm, black!80] (3.2,-.7) -- (3.2,.7);
        
            \fill[white, rotate=0] (1,-3mm) rectangle (2.2,1mm);
            \fill[white, rotate=0] (1,-3mm) rectangle (1.8,2mm);
        
            \draw[line width=0.5mm, shift={(1.3,0)}, rotate=-90] (-.8,.2) parabola bend (0,0) (.8,.2);
            \draw[line width=.3mm] (0.7,-2mm) -- (0.55,-2mm);
            \draw[line width=.3mm] (0.45,-2mm) -- (0.3,-2mm);
            \draw[line width=.3mm] (0.8,-2mm) -- (2,-2mm); 
            \draw[line width=.3mm] (2,0) -- (1.8,0); 
            \draw[line width=.3mm] (2,-2mm) arc (-90:90:1mm);
            \draw[line width=0.3mm, shift={(1.8,0)}, rotate=-90] (-.1,-.06) parabola bend (0,0) (.1,-.06);
    
            \node[fill=black!10] at (6,-1.6) {\color{black!60}\small far-field region};
            \node[] at (2,-1.6) {\color{black!60}\small near-field region};
    
            \node[] at (1.55,1.3) {\color{black}\small reflector antenna};
    
            \draw [black, line width=.15mm] plot [smooth,samples=200, tension=.8] coordinates { 
            (.65, 1.1) (.8, .7) (1.2, .3)};

        \end{scope}

    \end{tikzpicture}

    \caption{Qualitative sketch of the wavefronts of a parabolic reflector antenna (cf.~\cite[Fig.~2]{IEEE_standard_test_procedure_for_antennas}). 
    Close to the antenna, in its near-field region, the electromagnetic fields can accurately be approximated as plane waves; in its far-field region, the fields can accurately be approximated as spherical waves. 
    }
    \label{fig:misconscept_plane_spherical_wave}
\end{figure}
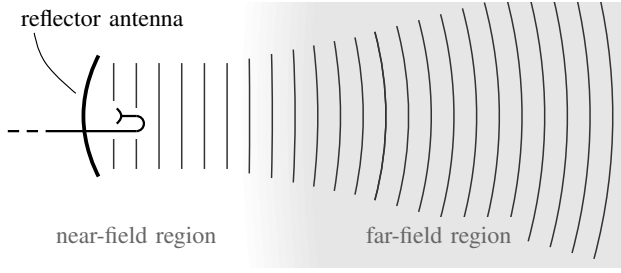

Furthermore, several recent publications state that, in the FF region, electromagnetic waves can be accurately approximated by a plane wave, whereas in the NF region, a spherical-wave model is required for accurate representation. 
This, again, is in disagreement with the IEEE definitions. 
\begin{rem}\label{rem:misconseption_plane_spherical_wave}
    In general, the NF and FF regions of an antenna system cannot be distinguished solely by examining whether the electromagnetic field at a given location can be accurately approximated as a plane wave or a spherical wave.
\end{rem} 
\begin{rem}
    It follows directly from~\fref{rem:misconseption_plane_spherical_wave} that the \emph{plane-wave assumption}---which states that,
    from the receiver's perspective, the incoming field can be approximated as a plane wave---is \emph{not}
    equivalent to the \emph{far-field assumption}---which states that, from the transmitter's perspective, the
    receiver lies in the far-field region (of the transmitter).
\end{rem}
To support~\fref{rem:misconseption_plane_spherical_wave}, consider an electrically large parabolic reflector antenna as illustrated in~\fref{fig:misconscept_plane_spherical_wave} that transmits a signal. 
Because of the parabolic geometry of the reflector, the waves leaving the antenna initially take the form of plane waves. 
Further away, due to \emph{beam divergence}, these plane waves begin to diverge and become spherical waves in the far-field region. 
At sufficiently large distances, the electromagnetic field can once again be accurately approximated, \emph{locally}, as a plane wave. 
In summary, as one moves away from this antenna, the fields can be represented first as a plane wave, second as a spherical wave, and then as both a plane and a spherical wave.\footnote{This paragraph is not intended as a formal proof of~\fref{rem:misconseption_plane_spherical_wave}, but rather to support the remark by invalidating the widespread misconception that, in a general antenna’s near-field region, a plane-wave model cannot accurately represent the electromagnetic fields.} 

Furthermore, several recent papers implicitly or explicitly claim that a higher signal frequency leads to a larger NF region. 
The authors of this paper assume that these claims result from inserting a smaller wavelength~$\lambda$ into the (Quasi-)Rayleigh\footnote{In~\fref{sec:rayleigh_1}, we detail why we refer to this distance as the \emph{Quasi}-Rayleigh distance.} distance formula we will provide in~\fref{eq:dist_qr} without adapting the largest dimension of the antenna system~$D_\up{S}$. 
This reasoning ignores the fact that antenna system dimensions are often scaled proportionally to the wavelength of interest.
In fact, when considering the common case scenario in which the largest dimension of an antenna system~$D_\up{S}$ is scaled proportionally with the wavelength~$\lambda$, equation~\fref{eq:dist_qr} implies that the near-field region actually shrinks at higher frequencies.

\begin{rem}\label{rem:frequency_size}
    Whether the extent of the NF region increases, remains unchanged, or decreases when an antenna system is designed and operated at a different frequency depends on the specific scenario. 
    In particular, it depends on whether the dimension of the antenna system is kept constant or scaled proportionally with the wavelength. 
\end{rem}

\section{Proposed Framework}
We now introduce our framework for \emph{analyzing} the field regions of antennas and \emph{testing} near-far-field boundaries.

\subsection{Idea of the Proposed Framework}\label{sec:idea}

It follows from the free-space Maxwell equations and the \emph{IEEE} far-field-region definition in~\cite[Sec.~4]{IEEE_standard_for_definitions_of_terms_for_antennas_2013} that \emph{in the far-field region} of an antenna system, the electromagnetic field radiated by this system at a certain frequency $f\in\mathbb{R}_{>0}$ can be accurately approximated as follows:\footnote{See, for example, Eq.~5.12 in~\cite{nieto_vesperinas_scattering_and_diffraction_in_physical_optics}.} 
\begin{align}
    \phv{E}(r,\theta,\varphi)
    &\approx
    \underbrace{
    Z_0^{\tfrac{1}{2}}
    \vect{f}(\theta,\varphi)\frac{e^{-jkr}}{r}
    }
    _{\triangleq\,\phv{E}_{\up{FF}}(r,\theta,\varphi)}
    \label{eq:far_field_E}
    \\
    \phv{H}(r,\theta,\varphi)
    &\approx
    \underbrace{
    Z_0^{-\tfrac{1}{2}}
    \!\big(
    \hat{\vect{r}}\times
    \vect{f}(\theta,\varphi)
    \big)
    \frac{e^{-jkr}}{r}}
    _{\triangleq\,\phv{H}_{\up{FF}}(r,\theta,\varphi)}.
    \label{eq:far_field_H}
\end{align}
Here, $k\triangleq 2\pi f\sqrt{\mu_\up{0}\varepsilon_\up{0}}$ is the free-space wavenumber, $\mu_\up{0}$ is the free-space permeability, $\varepsilon_\up{0}$ is the free-space permittivity, and $Z_0=\sqrt{\mu_\up{0}{\footnotesize/}\varepsilon_\up{0}}$ is the impedance of free space. 
Moreover, $\lambda\triangleq\frac{2\pi}{k}$ denotes the free-space wavelength.
Finally, $\vect{f}(\theta,\varphi)\in\mathbb{C}^3$ represents the \emph{angular field distribution}, which does not depend on the radial distance~$r$, and $\hat{\vect{r}}$ is the unit vector pointing in the direction~$(\theta,\varphi)$. 
\begin{defi}
    We use the physicist’s convention for spherical coordinate systems~\cite[Fig.~3]{ISO_quantities_and_units_2_mathematics}, with $r$ as the \emph{radial distance}, $\theta$ as the \emph{polar angle}, and $\varphi$ as the \emph{azimuthal angle}. 
\end{defi}

Based on the preceding discussion, it follows that a \emph{necessary} condition for a point $\vect{r}$, corresponding to $(r,\theta,\varphi)$, to lie in the far-field region of an antenna system is that the approximations in~\fref{eq:far_field_E} and~\fref{eq:far_field_H} are accurate\footnote{In~\fref{sec:metric}, we specify with which metric we examine, in this paper, whether an approximation is accurate or not.}; i.e. that it holds that $\phv{E}(\vect{r}) \approx \phv{E}_\up{FF}(\vect{r})$ and $\phv{H}(\vect{r}) \approx \phv{H}_\up{FF}(\vect{r})$.\footnote{The auxiliary field functions $\phv{E}_\up{FF}$ and $\phv{H}_\up{FF}$ are defined in~\fref{eq:far_field_E} and~\fref{eq:far_field_H}.} 
We emphasize that, for a given excitation, a \emph{sufficient} condition for a point to lie in the far-field region is that the approximations in~\fref{eq:far_field_E} and~\fref{eq:far_field_H} are also accurate for all points in the same direction that lie further away than that point. 
Specifically, for a system centered at the origin of the coordinate system, a point $\vect{r}$, corresponding to $(r,\theta,\varphi)$, lies in the far-field region if, for all radial distances $r' \geq r$, it holds that $\phv{E}(\vect{r}') \approx \phv{E}_\up{FF}(\vect{r}')$ and $\phv{H}(\vect{r}') \approx \phv{H}_\up{FF}(\vect{r}')$, where $\vect{r}'$ denotes the position corresponding to $(r',\theta,\varphi)$.
Thus, for a given direction $(\theta,\varphi)$, one can determine where the far-field region of an antenna system begins by analyzing how well the electromagnetic fields $\phv{E}$ and $\phv{H}$ match the auxiliary fields $\phv{E}_\up{FF}$ and $\phv{H}_\up{FF}$ as one increases the radial distance $r$ to the antenna system---this is the core idea underlying the test framework we propose next.

\subsection{Proposed Test Framework}\label{sec:framework}

The concept of the proposed test framework is illustrated in Figures~\ref{fig:concept_framework} and~\ref{fig:concept_example_plot}. 
Given an \emph{antenna system}, we first define the center of the coordinate system as a \emph{reference point}. 
This reference serves as the origin with respect to which the near-far-field boundaries will be evaluated.\footnote{For example, one can place the reference point on the surface of the antenna system, or, in the case of an aperture antenna, at the center of the aperture plane.}
Next, for a fixed observation direction $(\theta,\varphi)$, we define a \emph{test line} starting at the reference point and extending to infinity. 
For points $\vect{r}$ on this line, i.e., with polar angle $\theta$, azimuthal angle $\varphi$, and radial distance $r\geq0$, the electromagnetic field behavior will be analyzed. 
To accomplish this, the antenna system must be excited. 
For a single antenna element, the specific type of excitation is of minor importance as long as the antenna radiates a positive amount of power. 
In contrast, when dealing with an antenna array, a particular excitation scheme must be specified (e.g., beam steering toward a given point on the chosen test~line).\footnote{In our experiments, we apply the proposed framework to single-antenna scenarios, where beam steering is not applicable, and to antenna-array scenarios, where both NF and FF beamforming are used as excitation schemes, as detailed in~\fref{sec:test_scenarios}.}

Next, the auxiliary field functions, $\phv{E}_\up{FF}(\vect{r})$ and $\phv{H}_\up{FF}(\vect{r})$, for positions $\vect{r}$ on the test line must be obtained. 
To this end, we first calculate the angular field distribution vector $\vect{f}(\theta,\varphi)$ for the chosen direction by sampling either the electric field $\phv{E}(\vect{r}_\up{FF})$ or the magnetic field strength $\phv{H}(\vect{r}_\up{FF})$, as per Maxwell's equations, at a single point $\vect{r}_\up{FF}$ that must lie in the \emph{far-field region} and then use~\fref{eq:far_field_E} or~\fref{eq:far_field_H} to calculate $\vect{f}(\theta,\varphi)$ from the field at \emph{that} point $\vect{r}_\up{FF}$. 

\begin{rem}
    In general, when (analytically) calculating or measuring the electromagnetic fields, it is essential to select the point $\vect{r}_\up{FF}$ \emph{sufficiently} far from the antenna structure to ensure that it lies in the far-field region. 
    In case the antenna system is simulated using full-wave electromagnetic simulation software, such as Ansys HFSS~\cite{ansys_hffs}, the electromagnetic fields in the far-field region can be extracted directly, rendering an explicit choice of $\vect{r}_\up{FF}$ unnecessary.
\end{rem}

By using the definitions of $\phv{E}_\up{FF}(\vect{r})$ and $\phv{H}_\up{FF}(\vect{r})$ in~\fref{eq:far_field_E} and~\fref{eq:far_field_H} we can now calculate these auxiliary field functions for all positions $\vect{r}$ on the test line.  
Then, in a two-dimensional plot, such as the one sketched in~\fref{fig:concept_example_plot}, we illustrate how ``good'' the approximations in~\fref{eq:far_field_E} and~\fref{eq:far_field_H} are depending on the radial distance $r$.  
In other words, we analyze the approximation error between the (true) physical electromagnetic fields $\phv{E}(\vect{r})$ and $\phv{H}(\vect{r})$ and their auxiliary counterparts $\phv{E}_\up{FF}(\vect{r})$ and $\phv{H}_\up{FF}(\vect{r})$ along the test line. 
To quantify this ``approximation error,'' which we denote by $\epsilon_{(\theta,\varphi)}(r)$, one must choose a metric $\mu$ on the metric space $\mathbb{R}^3\times\mathbb{R}^3$.\footnote{The metric can be chosen to measure any application-specific performance.} 
The approximation error is then given by
\begin{align}
    \epsilon_{(\theta,\varphi)}(r)
    \triangleq
    \mu\Big(
    \phv{E}(\vect{r}),\phv{H}(\vect{r});\phv{E}_\up{FF}(\vect{r}),\phv{H}_\up{FF}(\vect{r})
    \Big),
\end{align}
where the position $\vect{r}$ corresponds to the spherical coordinates $(r,\theta,\varphi)$.
We detail our concrete choice of the metric $\mu$ used in this paper in~\fref{sec:metric}.
Finally, the near–far-field boundary under test is calculated and included in the plot as a vertical line. 
In order to interpret the performance of the near–far-field boundaries in the analyzed scenario, we make use of the aforementioned necessary and sufficient condition for a point to lie in the far-field region: 
(i)~If the approximation error $\epsilon_{(\theta,\varphi)}(r)$ is significantly greater than zero, then the point with spherical coordinates $(r,\theta,\varphi)$ does not lie in the far-field region of the antenna (system). 
(ii)~Given a point $\vect{r}$ with spherical coordinates $(r,\theta,\varphi)$. If, for all $r' \geq r$, it holds that $\epsilon_{(\theta,\varphi)}(r')$ remains small, then the point $\vect{r}$ lies in the far-field region of the antenna (system).

\subsection{Utilized Difference Metric}\label{sec:metric}
From comparing~\fref{eq:far_field_E} and~\fref{eq:far_field_H}, it follows that, in the far-field region, the electric field $\phv{E}$ is larger than the magnetic field strength $\phv{H}$ by the factor of the free-space impedance, $Z_0\approx\SI{377}{\Omega}$. 
Because we want to weight deviations of the electric and magnetic fields equally, we first normalize the electric field with $Z_0^{-1/2}$ and the magnetic field strength with $Z_0^{1/2}$, and then use the \emph{squared normalized Euclidean distance} as our difference metric:
\begin{align} \label{eq:differencemetric}
    &\mu\Big(
    \phv{E}(\vect{r}),\phv{H}(\vect{r});\phv{E}_\up{FF}(\vect{r}),\phv{H}_\up{FF}(\vect{r})
    \Big)
    \nonumber
    \\
    &\qquad \qquad \triangleq
    \left\{
    \begin{array}{ll}
    0
     & \text{\hspace{-3mm},\;\;if } \vect{F}=\vect{F}_\up{FF}=\vect{0} \\
        \left(
        \frac{
        \left\|
        \vect{F}
        -
        \vect{F}_\up{FF}
        \right\|_2
        }{
        \left\|
            \vect{F}
        \right\|_2
        +
        \left\|
            \vect{F}_\up{FF}
        \right\|_2
        }
        \right)^2\!
     & \text{\hspace{-3mm},\;\;otherwise.}
    \end{array}
    \right.
\end{align}
Here, we use
\begin{align}
    \vect{F}
    \triangleq
    \begin{bmatrix}
    Z_0^{-\frac{1}{2}}
    &\hspace{-4mm}\phv{E}(\vect{r})
    \\
    Z_0^{\frac{1}{2}}
    &\hspace{-4mm}\phv{H}(\vect{r})
    \end{bmatrix}
    \quad \textnormal{and} \quad
    \vect{F}_\up{FF}
    \triangleq
    \begin{bmatrix}
    Z_0^{-\frac{1}{2}}
    &\hspace{-4mm}\phv{E}_\up{FF}(\vect{r})
    \\
    Z_0^{\frac{1}{2}}
    &\hspace{-4mm}\phv{H}_\up{FF}(\vect{r})
    \end{bmatrix}.
\end{align}
The metric $\mu$ takes values between $0$ and $1$, where $\mu=0$ indicates that the compared fields are identical and $\mu=1$ indicates that they are antiparallel. 
Moreover, $\mu$ remains unchanged under scaling of all its input variables.
\begin{rem}
    We (somewhat arbitrarily) chose to use the difference metric $\mu$ in \fref{eq:differencemetric} to measure the approximation error because (i) it weights deviations of the electric and magnetic fields equally, (ii) it vanishes if and only if the fields are equal, and (iii) it is invariant under uniform scaling of all its input variables. 
    However, other difference metrics are also conceivable and may be more appropriate for certain applications.
\end{rem}


\section{Near-Far-Field Boundaries Under Test}

\begin{table}[]
\centering
\def\arraystretch{1.5}
\caption{Near–far-field boundaries included in our experiments. }\label{tab:distances_under_test}
\begin{tabular}{p{.6cm}p{3.4cm}lc}
\toprule
\multicolumn{2}{c}{Boundary Under Test} & \multicolumn{1}{c}{\hspace{-3mm}Reference} & \multicolumn{1}{c}{Eq.} \\
\midrule
$d_\up{QR}$&Quasi-Rayleigh Distance& \multicolumn{1}{c}{\hspace{-4mm}---} &\fref{eq:dist_qr}\\
$d_\up{AR}$&Direction-Dependent Rayleigh Distance\newline for Antenna Arrays&\cite[Sec. IV-B]{lu_zeng_communicating_with_extremely_large_array_surface}&\fref{eq:dist_ar}\\
$d_\up{UP}$&Uniform-Power Distance&\cite[Sec.~IV-A]{lu_zeng_communicating_with_extremely_large_array_surface}&\fref{eq:dist_up}\\
$d_\up{EN}$&Effective NF-Field\newline Boundary&\cite[Sec.~III-C2]{zhang_zhang_liu_di_codebook_design_and_beam_alignment_for_ios_aided_communications}&\fref{eq:dist_en}\\
$d_\up{EP}$&Equi-Power-Line Distance&\cite[Sec.~II]{renwang_shu_meixia_applicable_regions_of_spherical_and_plane_wave_modes_for_extremely_large_scale_array_communications}&\fref{eq:dist_ep}\\
$d_\up{WC}$&Optimal Worst-Case\newline Element-Wide Mismatch\newline Distance&\cite[Sec.~III]{daei_fodor_skoglund_when_near_becomes_far}&\fref{eq:dist_wc}\\\bottomrule
\end{tabular}
\end{table}

In the literature, various single-letter near-far-field boundary thresholds have been proposed to estimate where the NF region transitions into the FF region. 
In this work, we evaluate the performance of a \emph{selection} of existing near–far-field boundaries, as listed in~\fref{tab:distances_under_test}. 
In the following, we discuss these tested thresholds one by one.

\begin{rem}
    We selected the boundaries tested in this work in good faith to provide the reader with a representative overview of the types of boundaries proposed in the literature. 
    In~\fref{sec:excluded_distances}, we list several existing distance thresholds that were deliberately excluded from the experiments in this paper.
\end{rem}

\subsection{(Quasi-)Rayleigh Distance\label{sec:rayleigh_1}}
The concept of \emph{Rayleigh distance}\footnote{The term ``Rayleigh distance'' appears to have been introduced \emph{after} Rayleigh's time.}
(also sometimes referred to as the \emph{Fraunhofer distance}) is widely used in the wireless communications community to \emph{estimate}\footnote{We intentionally use the word ``estimate'' to highlight that the Rayleigh distance should \emph{not} be used to define where the FF region begins.} the location of the near–far-field boundary of an antenna system. 
However, in recent years, there have been instances where this concept has been applied to scenarios beyond its original intended scope and without proper justification. 
For this reason, we start by discussing the concept of Rayleigh distance. 
By the mid-20th century at the latest, the Rayleigh distance was used to refer to formulas that approximate the axial distance from an optical aperture---or an \emph{aperture antenna}---at which the difference in path lengths between the center and edges of the aperture to a point at that distance along the axis falls below a specified threshold.\footnote{The IEEE Standard~\cite[Sec.~1.6]{IEEE_standard_test_procedure_for_antennas} links the maximal path difference to the approximation formula; reference~\cite{wrixon_gain_measurements_of_standard_electromagnetic_horns} links the approximation formula to the term ``Rayleigh distance;''~and reference~\cite{bowhill_statistics_of_a_radio_wave_diffracted_by_a_random_ionosphere} confirms that the concept of Rayleigh distance was used both in antenna design and optics.}
A thorough introduction to aperture antennas, such as satellite dishes and horn antennas, can be found in the standard works of Balanis~\cite[Ch.~12]{balanis_antenna_theory_edition_3} and Stutzman and Thiele~\cite[Ch.~9]{stutzman_antenna_theory_and_design}. 
For a maximum path difference of $\lambda/16$, the Rayleigh distance of an aperture antenna is given as 
\begin{align}\label{eq:approximation_Rayleigh_distance_aperture_antennas}
    d_\up{R}\triangleq\frac{2 D_\up{P}^2}{\lambda},
\end{align}
where $D_\up{P}$ denotes the largest dimension of the antenna’s \emph{physical} aperture.\footnote{Depending on the source, the term ``Rayleigh distance'' may refer either (i) to the actual axial distance where the optical path-length difference falls below a chosen limit or (ii) to the corresponding approximation formula in the right-hand side of~\fref{eq:approximation_Rayleigh_distance_aperture_antennas}. For the sake of simplicity, in this paper, we define the Rayleigh distance as the corresponding approximation formula.}  
\begin{defi}[Physical Aperture]
    In this work, we use the term physical aperture of an \emph{aperture antenna} to refer to the finite planar surface through which it radiates and receives electromagnetic waves.
\end{defi}

\begin{rem}\label{rem:Dp_not_Ds}
    The largest dimension $D_\text{P}$ of an antenna’s physical aperture is generally not the same as the largest dimension~$D_\text{S}$ of the antenna itself. For example, many horn antennas have an elongated shape such that $D_\text{P} < D_\text{S}$. Moreover, for non-aperture antennas, the physical aperture is, in general, not well-defined. 
\end{rem}

As mentioned above, the right-hand side of~\fref{eq:approximation_Rayleigh_distance_aperture_antennas} is widely used to \emph{estimate} where the FF region of an \emph{aperture antenna} begins~\cite[Sec.~1.6]{IEEE_standard_test_procedure_for_antennas}.  

\begin{rem}\label{rem:rayleigh_defined_for_aperture_antennas}
The Rayleigh distance is defined for \emph{aperture antennas}. For other antenna types, particularly those without a well-defined physical aperture, the right-hand side of~\fref{eq:approximation_Rayleigh_distance_aperture_antennas} is, in general, not well-defined. 
\end{rem}
\begin{rem}
When generalizing the Rayleigh distance---or, more specifically, the right-hand side of~\fref{eq:approximation_Rayleigh_distance_aperture_antennas}---to antennas or antenna systems without a well-defined physical aperture, such as a dipole antenna, it is essential to clearly (and carefully) specify which substitute quantity is used in place of the largest dimension of the physical aperture $D_\up{P}$. 
\end{rem}

Many recent papers (sometimes carelessly) apply the right-hand side of~\fref{eq:approximation_Rayleigh_distance_aperture_antennas} to general antenna systems by substituting the largest dimension of the physical aperture~$D_\up{P}$ with the \emph{largest dimension of the antenna system}~$D_\up{S}$. 
In the following, we denote the resulting quasi-Rayleigh distance as
\begin{align}\label{eq:dist_qr}
    d_\up{QR}
    \triangleq
    \frac{2D_\up{S}^2}{\lambda}.
\end{align}

\begin{rem}\label{rem:do_not_use_pseudo_rayleigh_for_definition}
    To the best of our knowledge, there is no Maxwell-equation-based justification for interpreting the quasi-Rayleigh distance $d_\up{QR}$ as the boundary between the NF and FF regions. 
    This observation is further supported by the \emph{IEEE Standard Test Procedure}~\cite[Sec.~1.6]{IEEE_standard_test_procedure_for_antennas}, which explicitly states: ``For electrically-large antennas other than conventional broadside-aperture types, there is no recognized criterion defining the distance to the far-field inner boundary.'' 
\end{rem}


\subsection{Direction-Dependent Rayleigh Distance for Antenna Arrays}
In~\cite[Sec.~IV-B]{lu_zeng_communicating_with_extremely_large_array_surface}, Lu and Zeng propose a concept they refer to as the ``direction-dependent Rayleigh distance'' to ``extend the definition of the classical Rayleigh distance.'' 
However, since the definition in~\cite[Eq.~30]{lu_zeng_communicating_with_extremely_large_array_surface} is restricted to antenna arrays and cannot be applied to single antennas, we refer to their distance threshold in the following as the \emph{direction-dependent Rayleigh distance for antenna arrays}.
Given an array of $N$ antennas at positions $\{\vect{r}_n\}_{n\in[N]}$. 
Then, in a given direction $(\theta,\varphi)$, the direction-dependent Rayleigh distance for this antenna array is defined as~\cite[Eq.~30]{lu_zeng_communicating_with_extremely_large_array_surface}
\begin{align}\label{eq:dist_ar}
    d_\up{AR}(\theta,\varphi)
    \triangleq
    \inf \!\left\{r\in\mathbb{R}_{>0}\;\big|\;\Phi(r,\theta,\varphi)\leq \frac{\pi}{8}\right\}\!,
\end{align}
where $\Phi(r,\theta,\varphi)$ can be defined as~\cite[Eq.~29]{lu_zeng_communicating_with_extremely_large_array_surface}
\begin{align}
    \Phi(r,\theta,\varphi)
    \triangleq
    \max_{n\in[N]}
    k
    \Big(
        \|\vect{r}-\vect{r}_n\|_2
        -
        \|\vect{r}\|_2
        +
        \hat{\vect{r}}^\T 
        \vect{r}_n
    \Big).
\end{align}
Here, $\vect{r}$ is the (Cartesian) position vector corresponding to the position~$(r,\theta,\varphi)$ and  $\hat{\vect{r}}$ denotes the unit vector in the direction\mbox{$(\theta,\varphi)$}. 
Note that in the front direction, the direction-dependent Rayleigh distance for antenna arrays $d_\up{AR}$ is \emph{approximated} by the quasi-Rayleigh distance $d_\up{QR}$.

\subsection{Uniform-Power Distance}

In~\cite[Sec.~IV-A]{lu_zeng_communicating_with_extremely_large_array_surface}, 
Lu and Zeng propose another concept they refer to as the ``uniform-power distance.'' Given an array of $N$ antennas at positions $\{\vect{r}_n\}_{n\in[N]}$. 
Furthermore, let the antenna centers $\{\vect{r}_n\}_{n\in[N]}$ lie in a plane, and let $\hat{\vect{n}}$ be the unit vector normal to this plane, pointing in the direction in which the array shall transmit or receive signals.
Then, in a given direction $(\theta,\varphi)$, the uniform-power distance for this antenna array is defined as~\cite[Eq.~27]{lu_zeng_communicating_with_extremely_large_array_surface}
\begin{align}\label{eq:dist_up}
    d_\up{UP}(\theta,\varphi)
    \triangleq
    \inf \!\left\{r\in\mathbb{R}_{>0}\;\big|\;\Gamma(r,\theta,\varphi)\geq \Gamma_\up{th}\right\}\!,
\end{align}
where $\Gamma_\up{th}\in(0,1)$ is a certain threshold\footnote{In~\cite[Sec.~VI-B]{lu_zeng_communicating_with_extremely_large_array_surface}, a threshold value of $\Gamma_\up{th}=0.9$ is used.} 
and
\begin{align}
    \Gamma(r,\theta,\varphi)
    \triangleq
    \frac{\min_{n\in[N]}
    \frac{(\vect{r}-\vect{r}_n)^\T\hat{\vect{n}}}{\|\vect{r}-\vect{r}_n\|_2^3}
    }
    {\max_{n'\in[N]}
    \frac{(\vect{r}-\vect{r}_{n'})^\T\hat{\vect{n}}}{\|\vect{r}-\vect{r}_n'\|_2^3}
    }. 
\end{align}
Here, $\vect{r}$ is the (Cartesian) position vector corresponding to the position $(r,\theta,\varphi)$. 
%


%
\subsection{Effective NF-Field Boundary}

In~\cite[Sec.~III-C2]{zhang_zhang_liu_di_codebook_design_and_beam_alignment_for_ios_aided_communications}, 
Zhang \emph{et al.} propose the ``effective NF-field boundary.'' Given an array of $N$ antennas at the positions $\{\vect{r}_n\}_{n\in[N]}$. 
In a given direction $(\theta,\varphi)$, the effective NF-field boundary is given\footnote{In~\cite{zhang_zhang_liu_di_codebook_design_and_beam_alignment_for_ios_aided_communications}, it is not specified where the ``effective NF-field boundary'' should lie in the case where the function $r \mapsto \Psi(r,\theta,\varphi)$ crosses the threshold $\Psi_\up{th}$ multiple times. 
We hereby assume, in good faith, that it is the authors’ intention that, in this case, the last point that lies in the ``near-field region'' shall be taken as the boundary. 
Furthermore, the authors do not clearly specify what~$\vect{h}$ shall look like but instead refer to the sources~\cite{tang_wireless_communications_with_reconfigurable_intelligent_surface} and~\cite{zhou_spherical_wave_channel_and_analysis_for_large_linear_array_in_los_conditions}, from which we derive the form given in~\fref{eq:effective_nf_boundary_h}.} as~\cite[Eq.~10]{zhang_zhang_liu_di_codebook_design_and_beam_alignment_for_ios_aided_communications}
\begin{align}\label{eq:dist_en}
    d_\up{EN}(\theta,\varphi)
    \triangleq
    \sup \!\left\{r\in\mathbb{R}_{>0}\;\big|\;\Psi(r,\theta,\varphi)\geq \Psi_\up{th}\right\}\!,
\end{align}
where $\Psi_\up{th}\in \mathbb{R}_{\geq 0}$ is a certain threshold\footnote{In~\cite[Tab.~I]{zhang_zhang_liu_di_codebook_design_and_beam_alignment_for_ios_aided_communications}, a threshold value of $\Psi_\up{th}=1.05$ is used.}  
and
\begin{align}
    \Psi(r,\theta,\varphi)
    &\triangleq
    \frac{|\vect{h}^\T \vect{w}^\up{(NF)}|}{|\vect{h}^\T \vect{w}^\up{(FF)}|}
    \\
    \label{eq:effective_nf_boundary_h}
    \vect{h}
    &\triangleq
    \bigg[
    \frac{e^{-j k\|\vect{r}-\vect{r}_\up{1}\|_2}}{\|\vect{r}-\vect{r}_\up{1}\|_2}
    \cdots
    \frac{e^{-j k\|\vect{r}-\vect{r}_\up{N}\|_2}}{\|\vect{r}-\vect{r}_\up{N}\|_2}
    \bigg]^\T.
\end{align}
Here, $\vect{r}$ is the (Cartesian) position vector corresponding to the position $(r,\theta,\varphi)$, and $\vect{w}^\up{(FF)}$ and $\vect{w}^\up{(NF)}$ are defined in~\fref{eq:definition_w_FF} and~\fref{eq:definition_w_NF}, respectively. 


\subsection{Equi-Power-Line Distance}\label{sec:equi_power_line}

In~\cite[Sec.~II]{renwang_shu_meixia_applicable_regions_of_spherical_and_plane_wave_modes_for_extremely_large_scale_array_communications}, Renwang \emph{et al.} propose the ``equi-power line'' of a uniform linear array as a distance that indicates when a plane-wave model and when a spherical-wave model is more appropriate. 
The authors argue that a spherical-wave model is likely appropriate in the NF region, while a plane-wave model is more suitable in the FF region~\cite[Sec.~I]{renwang_shu_meixia_applicable_regions_of_spherical_and_plane_wave_modes_for_extremely_large_scale_array_communications}. 
Note that, in general, this claim cannot be justified, as we highlight in~\fref{rem:misconseption_plane_spherical_wave}. 
The authors of~\cite{renwang_shu_meixia_applicable_regions_of_spherical_and_plane_wave_modes_for_extremely_large_scale_array_communications} furthermore compare their equi-power-line distance to other distance thresholds, such as the ``classical Rayleigh distance.'' 
Consequently, we interpret the equi-power-line distance as an attempt to estimate the near–far-field boundary.
Given a uniform linear array of $N$ antennas at positions $\{\vect{r}_n\}_{n\in[N]}$. 
Then, in a given direction $(\theta,\varphi)$, the equi-power-line distance for this antenna array is given\footnote{In~\cite{renwang_shu_meixia_applicable_regions_of_spherical_and_plane_wave_modes_for_extremely_large_scale_array_communications}, it is not specified where the ``equi-power line'' should lie in the case where the function $r \mapsto \Upsilon(r,\theta,\varphi)$ crosses the threshold $\Upsilon_\up{th}$ multiple times. 
We hereby assume, in good faith, that it is the authors’ intention that, in this case, the last crossing point shall be taken as the boundary.}  as~\cite[Sec.~II]{renwang_shu_meixia_applicable_regions_of_spherical_and_plane_wave_modes_for_extremely_large_scale_array_communications}
\begin{align}\label{eq:dist_ep}
    d_\up{EP}(\theta,\varphi)
    \triangleq
    \sup \!\left\{r\in\mathbb{R}_{>0}\;\big|\;\Upsilon(r,\theta,\varphi)\leq \Upsilon_\up{th}\right\}\!,
\end{align}
where $\Upsilon_\up{th}\in (0,1)$ is a certain threshold\footnote{In~\cite[Sec.~II]{renwang_shu_meixia_applicable_regions_of_spherical_and_plane_wave_modes_for_extremely_large_scale_array_communications}, threshold values of $\Upsilon_\up{th}\in\{0.99,1.01\}$ are used.}   
and
\begin{align}
    \Upsilon(r,\theta,\varphi)
    \triangleq
    \frac{r^2}{N}\sum_{n\in [N]}\frac{1}{\|\vect{r}-\vect{r}_n\|_2^2}, 
\end{align}
where $\vect{r}$ is the (Cartesian) position vector corresponding to the position $(r,\theta,\varphi)$. 


\subsection{Optimal Worst-Case Element-Wide Mismatch Distance}
In~\cite[Sec.~III]{daei_fodor_skoglund_when_near_becomes_far}, for a uniform linear antenna array, Daei \emph{et al.} propose a ``worst-case element mismatch'' metric and use it to define what they call ``optimal distance.''\footnote{In~\cite{daei_fodor_skoglund_when_near_becomes_far}, the authors introduced multiple distances. For simplicity, we restrict our analysis to their first proposed distance.} 
Given a uniform linear array of $N$ antennas at positions $\{\vect{r}_n\}_{n\in[N]}$, 
the optimal worst-case element-wide mismatch distance can be defined as~\cite[Eq.~7]{daei_fodor_skoglund_when_near_becomes_far}
\begin{align}\label{eq:dist_wc}
    d_\up{WC}
    \triangleq
    \inf \!\left\{r\in\mathbb{R}_{>0}\;\big|\;\sup_{r'\geq r} \Xi(r')<\Xi_\up{th}\right\}\!,
\end{align}
where $\Xi_\up{th}\in \mathbb{R}_{\geq 0}$ is a certain threshold\footnote{In~\cite[Sec.~V]{daei_fodor_skoglund_when_near_becomes_far}, a threshold value of $\Xi_\up{th}=0.001$ is used.} 
and
\begin{align}
    \Xi(r)
    &\triangleq
    \max_{n\in [N]}
    \hspace{-1mm}
    \max_{
    \footnotesize
    \begin{matrix}\vect{a}\in\mathbb{R}^3\\:\!\|\vect{a}\|_2=1\end{matrix}}
    \Bigg|
    \frac{e^{-j k \|r\vect{a}-\vect{r}_n\|_2}}{\|r\vect{a}-\vect{r}_n\|_2}
    -
    \frac{e^{-j k (r-\vect{a}^\T\vect{r}_n)}}{r}
    \Bigg|
    .
\end{align}
%


\section{Excluded Distance Thresholds}\label{sec:excluded_distances}

We now briefly discuss several distance thresholds that we intentionally did not include in our experiments. 

\subsection{Rayleigh Distance}
We discussed the Rayleigh distance already in~\fref{sec:rayleigh_1}. 
In~\fref{rem:rayleigh_defined_for_aperture_antennas}, we highlight that the Rayleigh distance is defined only for \emph{aperture antennas}. 
Since our experiments do not involve aperture antennas, we did not test the Rayleigh distance but only the quasi-Rayleigh distance. 

\subsection{Björnson Distance}
In~\cite[Eq.~18]{bjoernson_a_primer_on_near_field_beamforming_for_arrays_and_reconfigurable_intelligent_surfaces}, Björnson \emph{et al.} introduce a distance threshold they call the ``Björnson distance.''
In recent papers of other authors, the Björnson distance has been listed erroneously as a near-far-field boundary. 
However, Björnson \emph{et al.} did \emph{not} claim that their distance serves this purpose. Therefore, we do not consider this concept further in this paper.  
\begin{rem}\label{rem:bjoernson_distance}
    The ``Bj\"ornson distance'' 
    introduced in~\cite[Eq.~18]{bjoernson_a_primer_on_near_field_beamforming_for_arrays_and_reconfigurable_intelligent_surfaces} is not a near-far-field boundary. 
    Rather, the authors interpret it in~\cite[Sec.~III-A]{bjoernson_a_primer_on_near_field_beamforming_for_arrays_and_reconfigurable_intelligent_surfaces} as the start of ``the array-equivalent of the Fresnel region.'' 
\end{rem}

\subsection{Distances Involving Multiple Antenna Systems}
In~\fref{sec:field_regions_intro}, we stated the definitions of the near-field and far-field regions of an antenna system according to the \emph{IEEE Standard for Definitions of Terms for Antennas}~\cite{IEEE_standard_test_procedure_for_antennas}. 
In particular, in~\fref{rem:intrinsic_property}, we emphasized that the field regions are an \emph{intrinsic} property of an antenna system and not the larger surrounding system in which they are embedded. 
Consequently, in our experiments, we do not consider distance thresholds that depend on both the transmitter and the receiver antenna systems, such as the ``MIMO'' distances proposed in~\cite{lu_dai_near_field_channel_estimation_in_mixed_los_nlos_environments_for_extremely_large_scale_mimo_systems} or the ``effective-degrees-of-freedom–based'' distance proposed in~\cite{sun_how_to_differentiate_between_near_far_field}.

\section{Evaluation of Near-Far-Field Boundaries}

We now describe the experiments conducted in this work. First, we detail the test scenarios used, and after that, we present the results of our experiments.

\subsection{Test Scenarios}\label{sec:test_scenarios}

\def\REMS at (#1,#2,#3){
    \draw  [shift={(#1,#2)},rotate=#3,line width=1pt, fill=white] (0,0) ellipse (0.3 and 0.3);
    \draw [shift={(#1,#2)},rotate=#3,line width=1pt] (-0.2,-0.05) -- (0.2,-0.05);
    \draw [shift={(#1,#2)},rotate=#3,line width=1pt, arrows = {-Stealth[inset=0, length=4pt, angle'=65]}] (0,-0.05) -- (0,0.25-0.05);
}

\def\Antenna at (#1,#2){
    \begin{scope}[canvas is xy plane at z=0]
        \begin{scope}[shift={(#1,#2)}]
            \draw[white, fill=white, line width=.3mm] (0,0) circle[radius=1.65mm];
            
            \draw[black, line width=.3mm] (0,0) circle[black, line width=.3pt, radius=1.5mm];
    
            \draw [fill=black, line width=0.01] (-0.07cm, -.075cm) rectangle (-0.04cm, .075cm);
    
            \draw [black, fill=black, line width=0.01] (-0.04cm, -.015) rectangle (0.05cm, .015);
    
            \draw [black, fill=black, line width=0.01] (0.01,0.05) -- (0.01,-0.05) -- (0.08,0);
        \end{scope}
    \end{scope}
}

\def\AntennaLarge at (#1,#2){
    \begin{scope}[canvas is xy plane at z=0]
        \begin{scope}[shift={(#1,#2)}, scale=7]
            \draw[white, fill=white, line width=7*.3mm] (0,0) circle[radius=1.65mm];
            
            \draw[black!20, line width=5*.3mm] (0,0) circle[black!20, line width=.3pt, radius=1.5mm];
    
            \draw [black!20, fill=black!20, line width=0.01] (-0.05cm, -.08cm) rectangle (-0.02cm, .08cm);
    
            \draw [black!20, fill=black!20, line width=0.01] (-0.04cm, -.015) rectangle (0.05cm, .015);
    
            \draw [black!20, fill=black!20, line width=0.01] (0.03,0.04) -- (0.03,-0.04) -- (0.1,0);
        \end{scope}
    \end{scope}
}

\begin{figure*}[htp!]
    \centering
    \subfloat[antenna arrangement in test scenarios]
    {
    \label{fig:scenario_sketch}
        {
        \small
        \tdplotsetmaincoords{60}{35}
        \begin{tikzpicture}[yscale=1,xscale=1]

            \clip (-3.5cm, -2.6cm) rectangle (2.6cm, 2.6cm);
        
           \begin{scope}[shift={(-1,0)}]
                \begin{scope}[tdplot_main_coords, scale=1.2]
        
                    \draw[line width=.2mm, black!100, arrows={-Triangle[angle=35:1.5mm]}] (-.5,0,0) -- (2.5,0,0) node[anchor=west] {$x$}; 
                    \draw[line width=.2mm, black!100, arrows={-Triangle[angle=35:1.5mm]}] (0,-2.5,0) -- (0,2.8,0) node[anchor=south west, shift={(0,-.2)}] {$y$}; 
                    \draw[line width=.2mm, black!100, arrows={-Triangle[angle=35:1.5mm]}] (0,0,-.5) -- (0,0,1.5) node[anchor=south] {$z$}; 
        
                    \draw[line width=.3mm] (0,0,0) -- (2,1,0);
                    \draw[line width=.3mm] (2*1.05,1*1.05,0) -- (2*1.15,1*1.15,0);
                    \draw[line width=.3mm] (2*1.2,1*1.2,0) -- (2*1.3,1*1.3,0);

                    \draw[line width=.3pt] (0,-2.1,0) --(-1.0,-2.1,0);
                    \draw[line width=.3pt] (0,-1.4,0) --(-1.0,-1.4,0);
                    \draw[line width=.3pt] (-.9,-1.3,0) --(-.9,-2.2,0);
                    \draw[line width=.3pt, arrows={-Triangle[angle=35:1.5mm]}] (-.9,-1.4,0) --(-.9,-2.1,0);
                    \draw[line width=.3pt, arrows={-Triangle[angle=35:1.5mm]}] (-.9,-2.1,0) -- (-.9,-1.4,0);
                    \node[anchor = south] at (-.9,-1.9,0) {\color{eth_black} $d$};

                    \draw[line width=.6pt, dotted] (0,-2.1,-.1) --(0,-2.1,.9);
                    \draw[line width=.6pt, dotted] (0,2.1,-.1) --(0,2.1,.9);

                    \draw[line width=.6pt, dotted] (0,-2.2,.7) --(0,-2.35,.7);
                    \draw[line width=.6pt, dotted] (0,2.2,.7) --(0,2.35,.7);

                    \Antenna at (0,0);
                    \Antenna at (0,.7);
                    \Antenna at (0,-.7);
                    \Antenna at (0,1.4);
                    \Antenna at (0,-1.4);
                    \Antenna at (0,2.1);
                    \Antenna at (0,-2.1);

                    \tdplotdrawarc{(0,0,0)}{1.8}{0}{27}{anchor=east, shift={(-0.02,0)}}{$\varphi$}
        
                    \draw[white, line width=1.2pt] (0,-.5,.7) -- (0,.5,.7);
                    \draw[line width=.6pt, dotted, arrows={-Triangle[angle=35:1.5mm]}] (0,0.015,.7) --(0,-2.1,.7);
                    \draw[line width=.6pt, dotted, arrows={-Triangle[angle=35:1.5mm]}] (0,-0.015,.7) -- (0,2.1,.7);
                    \node[anchor = south] at (0,-.5,.7) {\color{eth_black} $N$};

                \end{scope}
        
                \draw [black, line width=.15mm] plot [smooth,samples=200, tension=.8] coordinates { 
                (1.85, .35) (1.75, .2) (1.7,-.06)};
            
                \node[] at (2.45,.5) {\small test line};
        
                \node[anchor=west] at (-.8,-1.3) {\small infinitesimal dipole};
                \node[anchor=west] at (-.8,-1.7) {\small/ half-wave dipole antenna};
                \node[anchor=west] at (-.8,-2.1) {\small/ patch antenna};
                
               \draw [black, line width=.15mm] plot [smooth,samples=200, tension=.8] coordinates { 
                (0,-1.1) (-.4,-.85) (-.75,-.72)};

           \end{scope}
        \end{tikzpicture}
        }
    }
    \hspace{-5mm}
    \subfloat[infinitesimal dipole]
    {
    \label{fig:scenario_Hertz}
        {
        \small
        \tdplotsetmaincoords{60}{35}
        \begin{tikzpicture}[yscale=1,xscale=1]
            \clip (-1.5cm, -2.6cm) rectangle (1.5cm, 2.6cm);

            \begin{scope}[shift={(0,0)}, tdplot_main_coords]

                \draw [line width=1mm] (0,0,-.8) -- (0,0,0);
                
                \begin{scope}[canvas is xy plane at z=0]
                    \begin{scope}[shift={(0,0)}, scale=7]
                        \draw[opacity=.5, white, fill=white, line width=0] (0,0) circle[radius=1.5mm];
                        
                        \draw[black!20, line width=5*.3mm] (0,0) circle[black!20, line width=.3pt, radius=1.5mm];
                
                        \draw [black!20, fill=black!20, line width=0.01] (-0.05cm, -.08cm) rectangle (-0.02cm, .08cm);
                
                        \draw [black!20, fill=black!20, line width=0.01] (-0.04cm, -.015) rectangle (0.05cm, .015);
                
                        \draw [black!20, fill=black!20, line width=0.01] (0.03,0.04) -- (0.03,-0.04) -- (0.1,0);

                        \draw[fill=black, line width=0.001] (0,0) circle[radius=0.07mm];
                    \end{scope}
                \end{scope} 

                \draw [line width=1mm, arrows = {-Stealth[inset=0, length=4pt, angle'=65]}] (0,0,0) -- (0,0,.8);

            \end{scope}

        \end{tikzpicture}
        }
    }
    \hspace{-5mm}
    \subfloat[half-wave dipole antenna]
    {
    \label{fig:scenario_diploe}
        {
        \small
        \tdplotsetmaincoords{60}{35}
        \begin{tikzpicture}[yscale=1,xscale=1]
            \clip (-1.5cm, -2.6cm) rectangle (1.5cm, 2.6cm);

            \begin{scope}[shift={(0,0)}, tdplot_main_coords]

                \draw [line width=1mm] (0,0,-.8) -- (0,0,0);
                
                \begin{scope}[canvas is xy plane at z=0]
                    \begin{scope}[shift={(0,0)}, scale=7]
                        \draw[opacity=.5, white, fill=white, line width=0] (0,0) circle[radius=1.5mm];
                        
                        \draw[black!20, line width=5*.3mm] (0,0) circle[black!20, line width=.3pt, radius=1.5mm];
                
                        \draw [black!20, fill=black!20, line width=0.01] (-0.05cm, -.08cm) rectangle (-0.02cm, .08cm);
                
                        \draw [black!20, fill=black!20, line width=0.01] (-0.04cm, -.015) rectangle (0.05cm, .015);
                
                        \draw [black!20, fill=black!20, line width=0.01] (0.03,0.04) -- (0.03,-0.04) -- (0.1,0);

                    \end{scope}
                \end{scope} 
            \end{scope}

            \node[inner sep=0, anchor=center, xscale=-1] (image) at (0,0) {\includegraphics[width=.35\textwidth, trim={200 50 200 50}, clip]{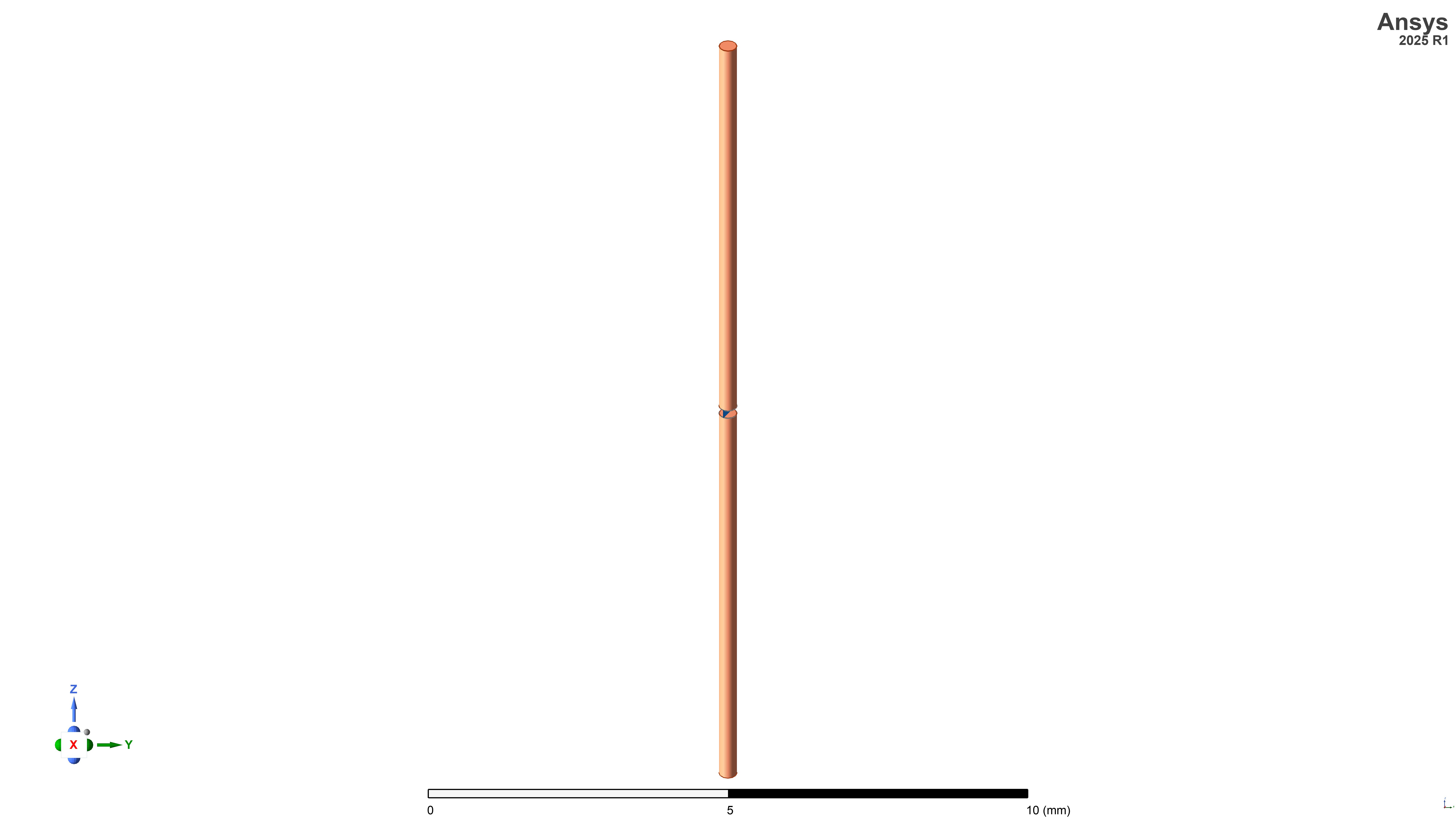}};
            
            \begin{scope}

            \clip (-5, -5) rectangle (5, 1);
            
            \begin{scope}[shift={(0,0)}, tdplot_main_coords]
                
                \begin{scope}[canvas is xy plane at z=0]
                    \begin{scope}[shift={(0,0)}, scale=7]
                        \draw[opacity=.5, white, fill=white, line width=0] (0,0) circle[radius=1.5mm];
                        
                        \draw[black!20, line width=5*.3mm] (0,0) circle[black!20, line width=.3pt, radius=1.5mm];
                
                        \draw [black!20, fill=black!20, line width=0.01] (-0.05cm, -.08cm) rectangle (-0.02cm, .08cm);
                
                        \draw [black!20, fill=black!20, line width=0.01] (-0.04cm, -.015) rectangle (0.05cm, .015);
                
                        \draw [black!20, fill=black!20, line width=0.01] (0.03,0.04) -- (0.03,-0.04) -- (0.1,0);

                    \end{scope}
                \end{scope} 
            \end{scope}
            \end{scope}

            \node[inner sep=0, anchor=center, xscale=-1] (image) at (0,0) {\includegraphics[width=.35\textwidth, trim={200 50 200 50}, clip]{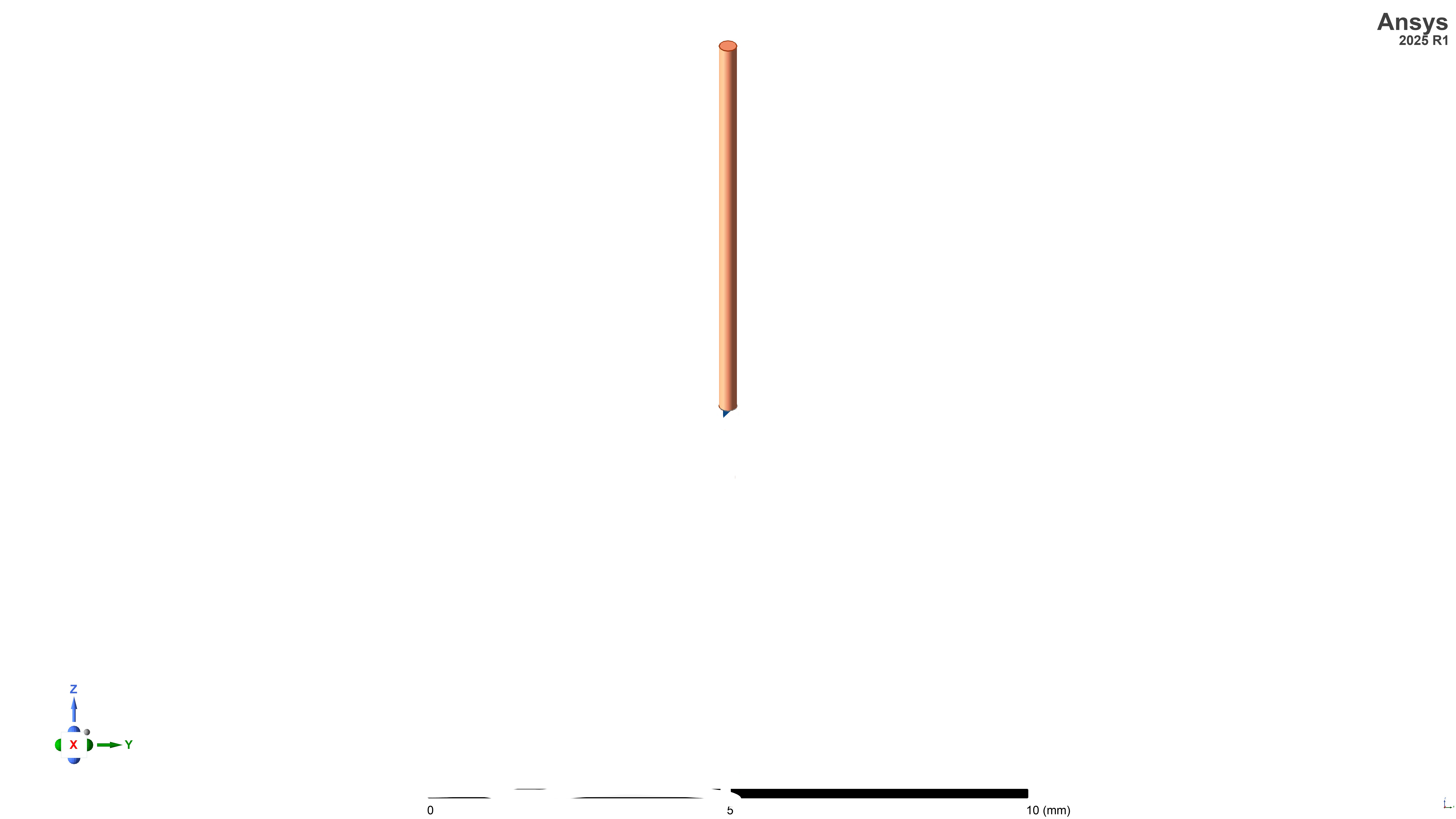}};

        \end{tikzpicture}
        }
    }
    \hspace{-5mm}
    \subfloat[patch antenna]
    {
    \label{fig:scenatio_patch}
        {
        \small
        \tdplotsetmaincoords{60}{35}
        \begin{tikzpicture}[yscale=1,xscale=1]
            \clip (-1.5cm, -2.6cm) rectangle (1.5cm, 2.6cm);

            \begin{scope}[shift={(0,0)}, tdplot_main_coords]

                \draw [line width=1mm] (0,0,-.8) -- (0,0,0);
                
                \begin{scope}[canvas is xy plane at z=0]
                    \begin{scope}[shift={(0,0)}, scale=7]
                        \draw[opacity=.5, white, fill=white, line width=0] (0,0) circle[radius=1.5mm];
                        
                        \draw[black!20, line width=5*.3mm] (0,0) circle[black!20, line width=.3pt, radius=1.5mm];
                
                        \draw [black!20, fill=black!20, line width=0.01] (-0.05cm, -.08cm) rectangle (-0.02cm, .08cm);
                
                        \draw [black!20, fill=black!20, line width=0.01] (-0.04cm, -.015) rectangle (0.05cm, .015);
                
                        \draw [black!20, fill=black!20, line width=0.01] (0.03,0.04) -- (0.03,-0.04) -- (0.1,0);

                    \end{scope}
                \end{scope} 
            \end{scope}

            \begin{scope}[shift={(-.12,+.07)}]
            \node[inner sep=0, anchor=center, xscale=-1] (image) at (0,0) {\includegraphics[width=.4\textwidth, trim={200 50 200 50}, clip]{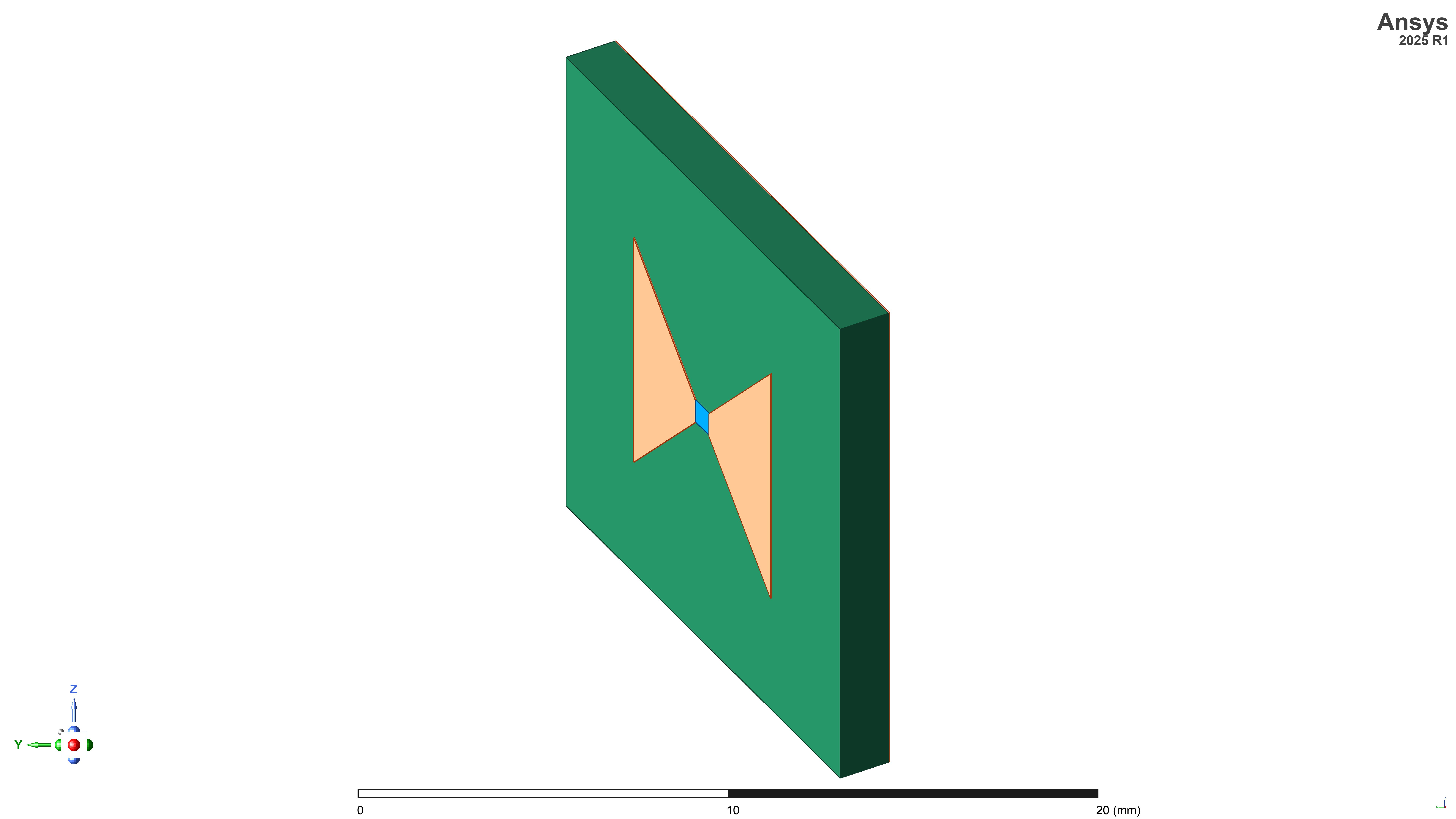}};
            \end{scope}
            
            \begin{scope}

            \clip (-5, -5) rectangle (5, 1);
            
            \begin{scope}[shift={(0,0)}, tdplot_main_coords]
                
                \begin{scope}[canvas is xy plane at z=0]
                    \begin{scope}[shift={(0,0)}, scale=7]
                        \draw[opacity=.5, white, fill=white, line width=0] (0,0) circle[radius=1.5mm];
                        
                        \draw[black!20, line width=5*.3mm] (0,0) circle[black!20, line width=.3pt, radius=1.5mm];
                
                        \draw [black!20, fill=black!20, line width=0.01] (-0.05cm, -.08cm) rectangle (-0.02cm, .08cm);
                
                        \draw [black!20, fill=black!20, line width=0.01] (-0.04cm, -.015) rectangle (0.05cm, .015);
                
                        \draw [black!20, fill=black!20, line width=0.01] (0.03,0.04) -- (0.03,-0.04) -- (0.1,0);

                    \end{scope}
                \end{scope} 
            \end{scope}
            \end{scope}

            \begin{scope}[shift={(-.12,+.07)}]
                \clip (-1.6,-1.65) -- (1.6,1.18) -- (1.6,4) -- (-1.6,4);
                \node[inner sep=0, anchor=center, xscale=-1] (image) at (0,0) {\includegraphics[width=.4\textwidth, trim={200 50 200 50}, clip]{figs/patch.png}};
            \end{scope}

        \end{tikzpicture}
        }
    }
\caption{Setup of the test scenarios used in the experiments. 
The analyzed antenna (array) lies on the $y$-axis, centered at the origin of coordinates, where we also placed the reference point of our framework. 
The $N$ array elements are uniformly separated by $d$, with each element being either an infinitesimal dipole (b), a half-wave dipole antenna (c), or a patch antenna (d). 
These three source types are shown in subfigures (b)--(d): subfigure (b) provides a qualitative sketch of the infinitesimal dipole aligned along the $z$-axis, while subfigures (c) and (d) contain screenshots taken from Ansys HFSS. 
The colors in the screenshots represent the following: orange for copper, green for FR4, and blue for lumped antenna ports. 
The half-wave dipole antenna has a length of $\lambda/2$. 
The patch antenna has a copper ground plane on its backside, and its side length is $\lambda/2$.}
\label{fig:scenarios}
\end{figure*}

For the experiments presented in this paper, we apply the framework described above to a range of antenna systems. 
Specifically, we consider uniform linear arrays, with the single-antenna setup included as a special case. 
The corresponding array configurations are illustrated in~\fref{fig:scenario_sketch}. 
In our experiments, we consider three types of antennas: infinitesimal dipole elements (see~\fref{fig:scenario_Hertz}), half-wave dipole antennas (see~\fref{fig:scenario_diploe}), and patch antennas (see~\fref{fig:scenatio_patch}).
For the infinitesimal dipole elements, the electromagnetic fields required in our framework can be obtained from closed-form expressions (see, e.g.,~\cite[Eq.~4.8 and 4.10]{balanis_antenna_theory_edition_3}).\footnote{Since an infinitesimal dipole element corresponds to the impulse whose response is described by the dyadic Green’s function, its electromagnetic fields can be directly calculated from the dyadic Green’s function.}
In our calculations, we assume that no inter-antenna coupling occurs between the infinitesimal dipole elements.\footnote{Consequently, the radiated fields of each dipole element are directly proportional to the corresponding excitation phasor.} 
For the other two antenna types, we obtained the electromagnetic fields through full-wave EM simulations in Ansys HFSS. 
In these simulations, the antennas are assumed to be excited by a signal source with an output impedance of \SI{50}{\ohm}, and inter-antenna coupling is inherently accounted for by the simulation software. 
In all experiments, the reference point was placed at the center of the arrays, i.e., the arrays were positioned at the origin of the coordinate system. 
The test line was located in the $xy$-plane ($\theta=90^\circ$) with its azimuthal angle chosen to be either (i) in the array’s front direction ($\varphi=0^\circ$), (ii) in the array's diagonal direction ($\varphi=45^\circ$), or (iii) in the array's side direction ($\varphi=90^\circ$). 
As excitation schemes, we use both FF beamforming (FF BF) in the direction of the test line and NF beamforming (NF BF) toward the respective points on the test line.\footnote{Specifically, FF BF corresponds to \emph{beamsteering} toward a specified direction, whereas NF BF corresponds to \emph{beamfocusing} toward a specific point.}
For FF BF, we consider an array of $N$ antennas located at positions $\{\vect{r}_n\}_{n\in[N]}$ and a test line oriented toward $(\theta,\varphi)$. 
The corresponding FF BF precoding vector is then given by 
\begin{align}\label{eq:definition_w_FF}
    \vect{w}^\up{(FF)}
    &\triangleq
    \big[
    e^{-j k
    \vect{r}_\up{1}^\T
    \hat{\vect{r}}
    }
    \cdots
    e^{-j k
    \vect{r}_\up{N}^\T
    \hat{\vect{r}}
    }
    \big]^\T. 
\end{align}
Here, $\hat{\vect{r}}$ denotes the unit vector pointing in the direction specified by $(\theta,\varphi)$. 
Consequently, the signal source corresponding to the $n$th antenna outputs a power wave (cf.~\cite[Eq.~1]{kurokawa_power_waves_and_scattering}) whose phasor is proportional to $\vect{w}^\up{(FF)}_n$.\footnote{We highlight that, in general, this beam-steering scheme does not maximize the power radiated toward the test-line direction. This is particularly the case when, due to coupling, the beam patterns of the individual antenna elements differ from one another.}  

For NF BF, we generalize the framework introduced in~\fref{sec:framework} by allowing different antenna excitations at each analyzed radial distance.\footnote{Consequently, for NF BF, the angular field distribution vector $\vect{f}$ must be obtained separately for each radial distance. Therefore, we restrict our consideration of NF BF to the infinitesimal dipole scenarios, where the required computations can be carried out quickly. }
Again, we consider an array of $N$ antennas located at positions $\{\vect{r}_n\}_{n\in[N]}$, and a test-line oriented toward $(\theta,\varphi)$. 
For a specific radial distance $r$, corresponding to the point $\vect{r}$ on the test line with coordinates $(r,\theta,\varphi)$, the NF BF precoding vector is given by\footnote{Again, we highlight that, in general, this beam-steering scheme does not maximize the power radiated toward the specified point on the test line.},
\begin{align}\label{eq:definition_w_NF}
    \vect{w}^\up{(NF)}
    &\triangleq
    \big[
    e^{j k
    \|\vect{r}-\vect{r}_\up{1}\|_2
    }
    \cdots
    e^{j k
    \|\vect{r}-\vect{r}_\up{N}\|_2
    }
    \big]^\T. 
\end{align}
We conclude this section with a brief explanation of why isotropic radiators were not considered in our experiments. 
An isotropic radiator is an idealized point source that radiates uniformly in all directions and polarizations. 
As highlighted in~\fref{rem:isotropic_radiator}, such antennas cannot exist. 
Moreover, following the proof in~\cite{mathis_a_short_proof_that_an_isotropic_antenna_is_impossible}, it becomes evident that no solution to Maxwell’s equations can exhibit the properties of an isotropic radiator. 
It is therefore not possible to apply our physically consistent framework to isotropic radiators. 
\begin{rem}\label{rem:isotropic_radiator}
    An isotropic radiator, in the sense of a coherent source radiating with uniform intensity in all directions and polarizations, cannot exist. 
    This impossibility follows directly from the definition of the far-field region, the free-space Maxwell's equations, and the \emph{Hairy Ball Theorem}, as demonstrated by Mathis~\cite{mathis_a_short_proof_that_an_isotropic_antenna_is_impossible}.
\end{rem}
Nonetheless, we emphasize that---especially for didactic purposes---it may still be useful to rely on the isotropic radiator model, especially when introducing fundamental concepts in wireless communications.

\subsection{Results}

\def\subfigWidth{8.9cm}
\def\subfigHeight{5cm}

\begin{figure*}[htp!]
    \subfloat
    {
        \begin{tikzpicture}
   
    \clip(-1.52cm,-.8cm) rectangle (\subfigWidth-1.52cm,\subfigHeight-.8cm);
    \begin{loglogaxis}[%
        title={$N\!=\!1$ antenna --- front direction},
        title style={at={(axis description cs:.5,.95)},anchor=south},
        width=6.75cm,
        height=3.45cm,
        scale only axis,
        xmin=1e-1,
        xmax=1e4,
        ymin=1e-4,
        ymax=1,
        xlabel={$r/\lambda$},
        x label style={at={(axis description cs:.5,-.09)},anchor=north},
        ylabel ={$\epsilon_{90^\circ\!,0^\circ}(r)$},
        grid=minor,
        tick style={line width=.5pt},
        ytick = {1, 1e-1, 1e-2, 1e-3, 1e-4, 1e-5, 1e-6},
        clip=false,
        minor ytick = {1e-6, 2e-6, 3e-6, 4e-6, 5e-6, 6e-6, 7e-6, 8e-6, 9e-6, 1e-5, 2e-5, 3e-5, 4e-5, 5e-5, 6e-5, 7e-5, 8e-5, 9e-5, 1e-4, 2e-4, 3e-4, 4e-4, 5e-4, 6e-4, 7e-4, 8e-4, 9e-4, 1e-3, 2e-3, 3e-3, 4e-3, 5e-3, 6e-3, 7e-3, 8e-3, 9e-3, 1e-2, 2e-2, 3e-2, 4e-2, 5e-2, 6e-2, 7e-2, 8e-2, 9e-2, 1e-1, 2e-1, 3e-1, 4e-1, 5e-1, 6e-1, 7e-1, 8e-1, 9e-1, 1, 2, 3, 4, 5},
        every outer y axis line/.append style={black,very thick},
        legend style={at={(0.97,0.97)},anchor=north east,legend cell align=left, align=left, draw=black,thick}
        ]

        \begin{scope}
            \clip (current axis.south west) rectangle (current axis.north east);
        
            \addplot[black!70, line width=1pt, dash pattern=on 7pt off 1pt on 1pt off 1pt] table [col sep=comma] {tikzfigs/data/d_inf_1_0.dat};
            
            \addplot[black!70, line width=1pt, dash pattern=on 6pt off 4pt] table [col sep=comma] {tikzfigs/data/d_2_1_0.dat};
            
            \addplot[black!70, line width=1pt, dash pattern=on 1pt off 1pt] table [col sep=comma] {tikzfigs/data/p_1_0.dat};

            \draw[opacity=0.9, blue, line width=1] (.5,1) -- (.5,1e-6);

            \draw[opacity=0.9, blue, line width=1] (1,1) -- (1,1e-6);
            
            \draw[white, fill=white] (.5,4e-2*4) circle (1.6mm);
            \draw[draw opacity=0.9, fill=white, draw=blue, line width=.7] (.5,4e-2*4) circle (1.4mm) node[] {\color{blue} \scriptsize 1};

            \draw[white, fill=white] (1,4e-2) circle (1.6mm);
            \draw[draw opacity=0.9, fill=white, draw=blue, line width=.7] (1,4e-2) circle (1.4mm) node[] {\color{blue} \scriptsize 1};

            \node[anchor=west] at (2.15,.17) {\color{blue}\small \contour{white}{half-wave dipole antenna}};

            \node[anchor=west] at (2.65,.019) {\color{blue}\small \contour{white}{patch antenna}};

            \draw [white, line width=.6mm] plot [smooth,samples=200, tension=.8] coordinates { 
            (.95,1.58e-1) (1.05,1.62e-1)};
            
            \draw [blue, line width=.15mm] plot [smooth,samples=200, tension=.8] coordinates { 
            (.72,1.5e-1) (1.5,1.75e-1) (2.3,1.8e-1)};

            \draw [blue, line width=.15mm] plot [smooth,samples=200, tension=.8] coordinates { 
            (1.4,0.031) (2,0.022) (2.7,0.18e-1)};

        \end{scope}

        \draw[black, line width=1pt] (current axis.south west) rectangle (current axis.north east);
       
    \end{loglogaxis}
\end{tikzpicture}   
    }
    \hspace{-4mm}
    \subfloat
    {
        \begin{tikzpicture}
    \clip(-1.5cm,-5mm) rectangle (\subfigWidth-1.5cm,\subfigHeight-5mm);
    \begin{scope}[shift={(0,0)}]
    
        \begin{scope}[shift={(0cm,3.7cm)}];   
            \draw[black!70, line width=1pt, dash pattern=on 7pt off 1pt on 1pt off 1pt] (0,0) -- (25pt,0);
            \node[anchor=west] at (26pt,0) {infinitesimal dipole sources (FF BF)};
        \end{scope}

        \begin{scope}[shift={(0cm,3.3cm)}];   
            \draw[black!70, line width=1pt, dash pattern=on 5pt off 1pt on 1pt off 1pt on 1pt off 1pt] (0,0) -- (25pt,0);
            \node[anchor=west] at (26pt,0) {infinitesimal dipole sources (NF BF)};
        \end{scope}

        \begin{scope}[shift={(0cm,2.9cm)}];   
            \draw[black!70, line width=1pt, dash pattern=on 6pt off 4pt] (0,0) -- (25pt,0);
            \node[anchor=west] at (26pt,0) {half-wave dipole antennas (FF BF)};
        \end{scope}

        \begin{scope}[shift={(0cm,2.5cm)}];   
            \draw[black!70, line width=1pt, dash pattern=on 1pt off 1pt] (0,0) -- (25pt,0);
            \node[anchor=west] at (26pt,0) {patch antennas (FF BF)};
        \end{scope}
    
        \begin{scope}[shift={(0cm,2cm)}];   
            \draw[opacity=0.9, blue, line width=1] (0,0) -- (25pt,0);
            \draw[white, fill=white] (12.5pt,0) circle (1.6mm);
            \draw[draw opacity=0.9, fill=black!0, draw=blue, line width=.7] (12.5pt,0) circle (1.4mm) node[] {\color{blue} \scriptsize 1};
            \node[anchor=west] at (26pt,0) {$d_\up{QR}$};
        
            \begin{scope}[shift={(3.5cm,0)}]
                \draw[opacity=0.9, bronze_80, line width=1] (0,0) -- (25pt,0);
                \draw[white, fill=white] (12.5pt,0) circle (1.6mm);
                \draw[draw opacity=0.9, fill=black!0, draw=bronze_80, line width=.7] (12.5pt,0) circle (1.4mm) node[] {\color{bronze_80} \scriptsize \hspace{.3mm}4\hspace{-.2mm}'};
                \node[anchor=west] at (26pt,0) {$d_\up{EN}$};
                \node[anchor=west] at (46pt,0) {\footnotesize $(\Psi_\up{th}=1.05)$};
            \end{scope}
        \end{scope}
    
        \begin{scope}[shift={(0cm,1.6cm)}];  
            \draw[opacity=0.9, petrol, line width=1] (0,0) -- (25pt,0);
            \draw[white, fill=white] (12.5pt,0) circle (1.6mm);
            \draw[draw opacity=0.9, fill=black!0, draw=petrol, line width=.7] (12.5pt,0) circle (1.4mm) node[] {\color{petrol} \scriptsize 2};
            \node[anchor=west] at (26pt,0) {$d_\up{AR}$};
        
            \begin{scope}[shift={(3.5cm,0)}]
                \draw[opacity=0.9, purple_120, line width=1] (0,0) -- (25pt,0);
                \draw[white, fill=white] (12.5pt,0) circle (1.6mm);
                \draw[draw opacity=0.9, fill=black!0, draw=purple_120, line width=.7] (12.5pt,0) circle (1.4mm) node[] {\color{purple_120} \scriptsize 5};
                \node[anchor=west] at (26pt,0) {$d_\up{EP}$};
                \node[anchor=west] at (46pt,0) {\footnotesize $(\Upsilon_\up{th}=1.01)$};
            \end{scope}
        \end{scope}
    
        \begin{scope}[shift={(0cm,1.2cm)}];  
            \draw[opacity=0.9, green_120, line width=1] (0,0) -- (25pt,0);
            \draw[white, fill=white] (12.5pt,0) circle (1.6mm);
            \draw[draw opacity=0.9, fill=black!0, draw=green_120, line width=.7] (12.5pt,0) circle (1.4mm) node[] {\color{green_120} \scriptsize 3};
            \node[anchor=west] at (26pt,0) {$d_\up{UP}$};
            \node[anchor=west] at (46pt,0) {\footnotesize $(\Gamma_\up{th}=0.9)$};
            
            \begin{scope}[shift={(3.5cm,0)}]
                \draw[opacity=0.9, purple_80, line width=1] (0,0) -- (25pt,0);
                \draw[white, fill=white] (12.5pt,0) circle (1.6mm);
                \draw[draw opacity=0.9, fill=black!0, draw=purple_80, line width=.7] (12.5pt,0) circle (1.4mm) node[] {\color{purple_80} \scriptsize \hspace{.3mm}5\hspace{-.2mm}'};
                \node[anchor=west] at (26pt,0) {$d_\up{EP}$};
                \node[anchor=west] at (46pt,0) {\footnotesize $(\Upsilon_\up{th}=0.99)$};
            \end{scope}
        \end{scope}
    
        \begin{scope}[shift={(0cm,0.8cm)}];  
            \draw[opacity=0.9, green_80, line width=1, dash pattern=on 8.5pt off 1pt on 2pt off 1pt on 1pt off 1pt on 8.5 off 0] (0,0) -- (25pt,0);
            \draw[white, fill=white] (12.5pt,0) circle (1.6mm);
            \draw[draw opacity=0.9, fill=black!0, draw=green_80, line width=.7] (12.5pt,0) circle (1.4mm) node[] {\color{green_80} \scriptsize \hspace{.3mm}3\hspace{-.2mm}'};
            \node[anchor=west] at (26pt,0) {$d_\up{UP}$};
            \node[anchor=west] at (46pt,0) {\footnotesize $(\Gamma_\up{th}=0.8)$};
        
            \begin{scope}[shift={(3.5cm,0)}]
                \draw[opacity=0.9, red_120, line width=1] (0,0) -- (25pt,0);
                \draw[white, fill=white] (12.5pt,0) circle (1.6mm);
                \draw[draw opacity=0.9, fill=black!0, draw=red_120, line width=.7] (12.5pt,0) circle (1.4mm) node[] {\color{red_120} \scriptsize 6};
                \node[anchor=west] at (26pt,0) {$d_\up{WC}$};
                \node[anchor=west] at (46pt,0) {\footnotesize $(\Xi_\up{th}=0.001)$};
            \end{scope}
        \end{scope}
    
        \begin{scope}[shift={(0cm,0.4cm)}];   
            \draw[opacity=0.9, bronze_120, line width=1] (0,0) -- (25pt,0);
            \draw[white, fill=white] (12.5pt,0) circle (1.6mm);
            \draw[draw opacity=0.9, fill=black!0, draw=bronze_120, line width=.7] (12.5pt,0) circle (1.4mm) node[] {\color{bronze_120} \scriptsize 4};
            \node[anchor=west] at (26pt,0) {$d_\up{EN}$};
            \node[anchor=west] at (46pt,0) {\footnotesize $(\Psi_\up{th}=1.01)$};
        
            \begin{scope}[shift={(3.5cm,0)}]
                \draw[opacity=0.9, red_80, line width=1] (0,0) -- (25pt,0);
                \draw[white, fill=white] (12.5pt,0) circle (1.6mm);
                \draw[draw opacity=0.9, fill=black!0, draw=red_80, line width=.7] (12.5pt,0) circle (1.4mm) node[] {\color{red_80} \scriptsize \hspace{.3mm}6\hspace{-.2mm}'};
                \node[anchor=west] at (26pt,0) {$d_\up{WC}$};
                \node[anchor=west] at (46pt,0) {\footnotesize $(\Xi_\up{th}=0.01)$};
            \end{scope}
        \end{scope}
    \end{scope}
\end{tikzpicture}   
    }
    \\[-0.32cm]
    \subfloat
    {
        \begin{tikzpicture}
   
    \clip(-1.52cm,-.8cm) rectangle (\subfigWidth-1.52cm,\subfigHeight-.8cm);
    \begin{loglogaxis}[%
        title={$N\!=\!8$ antennas --- front direction},
        title style={at={(axis description cs:.5,.95)},anchor=south},
        width=6.75cm,
        height=3.45cm,
        scale only axis,
        xmin=1e-1,
        xmax=1e4,
        ymin=1e-4,
        ymax=1,
        xlabel={$r/\lambda$},
        x label style={at={(axis description cs:.5,-.09)},anchor=north},
        ylabel ={$\epsilon_{90^\circ\!,0^\circ}(r)$},
        grid=minor,
        tick style={line width=.5pt},
        ytick = {1, 1e-1, 1e-2, 1e-3, 1e-4, 1e-6},
        clip=false,
        minor ytick = {1e-6, 2e-6, 3e-6, 4e-6, 5e-6, 6e-6, 7e-6, 8e-6, 9e-6, 1e-5, 2e-5, 3e-5, 4e-5, 5e-5, 6e-5, 7e-5, 8e-5, 9e-5, 1e-4, 2e-4, 3e-4, 4e-4, 5e-4, 6e-4, 7e-4, 8e-4, 9e-4, 1e-3, 2e-3, 3e-3, 4e-3, 5e-3, 6e-3, 7e-3, 8e-3, 9e-3, 1e-2, 2e-2, 3e-2, 4e-2, 5e-2, 6e-2, 7e-2, 8e-2, 9e-2, 1e-1, 2e-1, 3e-1, 4e-1, 5e-1, 6e-1, 7e-1, 8e-1, 9e-1, 1, 2, 3, 4, 5},
        every outer y axis line/.append style={black,very thick},
        legend style={at={(0.97,0.97)},anchor=north east,legend cell align=left, align=left, draw=black,thick}
        ]

        \begin{scope}
            \clip (current axis.south west) rectangle (current axis.north east);

            \addplot[black!70, line width=1pt, dash pattern=on 7pt off 1pt on 1pt off 1pt] table [col sep=comma] {tikzfigs/data/d_inf_8_0_ff.dat};

            \addplot[black!70, line width=1pt, dash pattern=on 5pt off 1pt on 1pt off 1pt on 1pt off 1pt] table [col sep=comma] {tikzfigs/data/d_inf_8_0_nf.dat};
            
            \addplot[black!70, line width=1pt, dash pattern=on 6pt off 4pt] table [col sep=comma] {tikzfigs/data/d_2_8_0.dat};
            
            \addplot[black!70, line width=1pt, dash pattern=on 1pt off 1pt] table [col sep=comma] {tikzfigs/data/p_8_0.dat};

            \draw[opacity=0.9, blue, line width=1] (24.5,1) -- (24.5,1e-6);

            \draw[opacity=0.9, petrol, line width=1] (24.6827,1) -- (24.6827,1e-6);

            \draw[opacity=0.9, green_120, line width=1] (6.4835,1) -- (6.4835,1e-6);
           
            \draw[opacity=0.9, green_80, line width=1] (4.3315,1) -- (4.3315,1e-6);

            \draw[opacity=0.9, bronze_120, line width=1] (25.2582,1) -- (25.2582,1e-6);

            \draw[opacity=0.9, bronze_80, line width=1] (11.4040,1) -- (11.4040,1e-6);
          
            \draw[opacity=0.9, purple_120, line width=1] (1000,1) -- (1000,1e-6);
           
            \draw[opacity=0.9, purple_80, line width=1] (11.2733,1) -- (11.2733,1e-6);

            \draw[opacity=0.9, red_120, line width=1] (560.7235,1) -- (560.7235,1e-6);
            
            \draw[opacity=0.9, red_80, line width=1] (177.1121,1) -- (177.1121,1e-6);

            \draw[white, fill=white] (24.5,2e-4*9) circle (1.6mm);
            \draw[draw opacity=0.9, fill=white, draw=blue, line width=.7] (24.5,2e-4*9) circle (1.4mm) node[] {\color{blue} \scriptsize 1};
            
            \draw[white, fill=white] (24.6827,2e-4*3) circle (1.6mm);
            \draw[draw opacity=0.9, fill=white, draw=petrol, line width=.7] (24.6827,2e-4*3) circle (1.4mm) node[] {\color{petrol} \scriptsize 2};
            
            \draw[white, fill=white] (6.4835,1e-1/3) circle (1.6mm);
            \draw[draw opacity=0.9, fill=white, draw=green_120, line width=.7] (6.4835,1e-1/3) circle (1.4mm) node[] {\color{green_120} \scriptsize 3};

            \draw[white, fill=white] (4.3315,2.5e-1/3) circle (1.6mm);
            \draw[draw opacity=0.9, fill=white, draw=green_80, line width=.7] (4.3315,2.5e-1/3) circle (1.4mm) node[] {\color{green_80} \scriptsize \hspace{.3mm}3\hspace{-.2mm}'};

            \draw[white, fill=white] (25.2582,2e-4) circle (1.6mm);
            \draw[draw opacity=0.9, fill=white, draw=bronze_120, line width=.7] (25.2582,2e-4) circle (1.4mm) node[] {\color{bronze_120} \scriptsize 4};

            \draw[white, fill=white] (11.4040,4e-2/3) circle (1.6mm);
            \draw[draw opacity=0.9, fill=white, draw=bronze_80, line width=.7] (11.4040,4e-2/3) circle (1.4mm) node[] {\color{bronze_80} \scriptsize \hspace{.3mm}4\hspace{-.2mm}'};
 
            \draw[white, fill=white] (1000,2e-4) circle (1.6mm);
            \draw[draw opacity=0.9, fill=white, draw=purple_120, line width=.7] (1000,2e-4) circle (1.4mm) node[] {\color{purple_120} \scriptsize 5};
    
            \draw[white, fill=white] (11.2733,4e-2/3/3) circle (1.6mm);
            \draw[draw opacity=0.9, fill=white, draw=purple_80, line width=.7] (11.2733,4e-2/3/3) circle (1.4mm) node[] {\color{purple_80} \scriptsize \hspace{.3mm}5\hspace{-.2mm}'};
    
            \draw[white, fill=white] (560.7235,2e-4) circle (1.6mm);
            \draw[draw opacity=0.9, fill=white, draw=red_120, line width=.7] (560.7235,2e-4) circle (1.4mm) node[] {\color{red_120} \scriptsize 6};
            
            \draw[white, fill=white] (177.1121,2e-4*3) circle (1.6mm);
            \draw[draw opacity=0.9, fill=white, draw=red_80, line width=.7] (177.1121,2e-4*3) circle (1.4mm) node[] {\color{red_80} \scriptsize \hspace{.3mm}6\hspace{-.2mm}'};

        \end{scope}

        \draw[black, line width=1pt] (current axis.south west) rectangle (current axis.north east);
       
    \end{loglogaxis}
\end{tikzpicture}   
    }
    \hspace{-4mm}
    \subfloat
    {
        \begin{tikzpicture}
   
    \clip(-1.52cm,-.8cm) rectangle (\subfigWidth-1.52cm,\subfigHeight-.8cm);
    \begin{loglogaxis}[%
        title={$N\!=\!64$ antennas --- front direction},
        title style={at={(axis description cs:.5,.95)},anchor=south},
        width=6.75cm,
        height=3.45cm,
        scale only axis,
        xmin=1e-1,
        xmax=1e4,
        ymin=1e-4,
        ymax=1,
        xlabel={$r/\lambda$},
        x label style={at={(axis description cs:.5,-.09)},anchor=north},
        ylabel ={$\epsilon_{90^\circ\!,0^\circ}(r)$},
        grid=minor,
        tick style={line width=.5pt},
        ytick = {1, 1e-1, 1e-2, 1e-3, 1e-4, 1e-6},
        clip=false,
        minor ytick = {1e-6, 2e-6, 3e-6, 4e-6, 5e-6, 6e-6, 7e-6, 8e-6, 9e-6, 1e-5, 2e-5, 3e-5, 4e-5, 5e-5, 6e-5, 7e-5, 8e-5, 9e-5, 1e-4, 2e-4, 3e-4, 4e-4, 5e-4, 6e-4, 7e-4, 8e-4, 9e-4, 1e-3, 2e-3, 3e-3, 4e-3, 5e-3, 6e-3, 7e-3, 8e-3, 9e-3, 1e-2, 2e-2, 3e-2, 4e-2, 5e-2, 6e-2, 7e-2, 8e-2, 9e-2, 1e-1, 2e-1, 3e-1, 4e-1, 5e-1, 6e-1, 7e-1, 8e-1, 9e-1, 1, 2, 3, 4, 5},
        every outer y axis line/.append style={black,very thick},
        legend style={at={(0.97,0.97)},anchor=north east,legend cell align=left, align=left, draw=black,thick}
        ]

        \begin{scope}
            \clip (current axis.south west) rectangle (current axis.north east);
        
            \addplot[black!70, line width=1pt, dash pattern=on 7pt off 1pt on 1pt off 1pt] table [col sep=comma] {tikzfigs/data/d_inf_64_0_ff.dat};

            \addplot[black!70, line width=1pt, dash pattern=on 5pt off 1pt on 1pt off 1pt on 1pt off 1pt] table [col sep=comma] {tikzfigs/data/d_inf_64_0_nf.dat};
            
            \addplot[black!70, line width=1pt, dash pattern=on 6pt off 4pt] table [col sep=comma] {tikzfigs/data/d_2_64_0.dat};
            
            \addplot[black!70, line width=1pt, dash pattern=on 1pt off 1pt] table [col sep=comma] {tikzfigs/data/p_64_0.dat};

            \draw[opacity=0.9, blue, line width=1] (1.9845e+3,1) -- (1.9845e+3,1e-6);

            \draw[opacity=0.9, petrol, line width=1] (1.9920e+3,1) -- (1.9920e+3,1e-6);

            \draw[opacity=0.9, green_120, line width=1] (58.5825,1) -- (58.5825,1e-6);
           
            \draw[opacity=0.9, green_80, line width=1] (39.5911,1) -- (39.5911,1e-6);

            \draw[opacity=0.9, bronze_120, line width=1] (1.6952e+3,1) -- (1.6952e+3,1e-6);

            \draw[opacity=0.9, bronze_80, line width=1] (765.3919,1) -- (765.3919,1e-6);
          
           
            \draw[opacity=0.9, purple_80, line width=1] (90.7733,1) -- (90.7733,1e-6);

            \draw[opacity=0.9, red_120, line width=1] (5.0665e+3,1e-4) -- (5.0665e+3,1.5e-1);
            \draw[opacity=0.9, red_120, line width=1] (5.0665e+3,1.85e-1) -- (5.0665e+3,1);
            
            \draw[opacity=0.9, red_80, line width=1] (1.6003e+3,1) -- (1.6003e+3,1e-6);

            \draw[white, fill=white] (1.9845e+3,4e-3*9) circle (1.6mm);
            \draw[draw opacity=0.9, fill=white, draw=blue, line width=.7] (1.9845e+3,4e-3*9) circle (1.4mm) node[] {\color{blue} \scriptsize 1};
            
            \draw[white, fill=white] (1.9920e+3,4e-3*3) circle (1.6mm);
            \draw[draw opacity=0.9, fill=white, draw=petrol, line width=.7] (1.9920e+3,4e-3*3) circle (1.4mm) node[] {\color{petrol} \scriptsize 2};
            
            \draw[white, fill=white] (58.5825,5e-1/3) circle (1.6mm);
            \draw[draw opacity=0.9, fill=white, draw=green_120, line width=.7] (58.5825,5e-1/3) circle (1.4mm) node[] {\color{green_120} \scriptsize 3};

            \draw[white, fill=white] (39.5911,5e-1/3/3) circle (1.6mm);
            \draw[draw opacity=0.9, fill=white, draw=green_80, line width=.7] (39.5911,5e-1/3/3) circle (1.4mm) node[] {\color{green_80} \scriptsize \hspace{.3mm}3\hspace{-.2mm}'};

            \draw[white, fill=white] (1.6952e+3,6e-3/3) circle (1.6mm);
            \draw[draw opacity=0.9, fill=white, draw=bronze_120, line width=.7] (1.6952e+3,6e-3/3) circle (1.4mm) node[] {\color{bronze_120} \scriptsize 4};

            \draw[white, fill=white] (765.3919,3e-2/3) circle (1.6mm);
            \draw[draw opacity=0.9, fill=white, draw=bronze_80, line width=.7] (765.3919,3e-2/3) circle (1.4mm) node[] {\color{bronze_80} \scriptsize \hspace{.3mm}4\hspace{-.2mm}'};
 
            \draw [purple_120, line width=.8, arrows = {-Stealth[inset=0, length=4pt, angle'=65]}] (4000,5e-1) -- (4000,5e-1/3) -- (9000,5e-1/3);
            \draw[white, fill=white] (4000,5e-1) circle (1.6mm);
            \draw[draw opacity=0.9, fill=white, draw=purple_120, line width=.7] (4000,5e-1) circle (1.4mm) node[] {\color{purple_120} \scriptsize 5};
    
            \draw[white, fill=white] (90.7733,5e-1/3/3) circle (1.6mm);
            \draw[draw opacity=0.9, fill=white, draw=purple_80, line width=.7] (90.7733,5e-1/3/3) circle (1.4mm) node[] {\color{purple_80} \scriptsize \hspace{.3mm}5\hspace{-.2mm}'};
    
            \draw[white, fill=white] (5.0665e+3,2e-4) circle (1.6mm);
            \draw[draw opacity=0.9, fill=white, draw=red_120, line width=.7] (5.0665e+3,2e-4) circle (1.4mm) node[] {\color{red_120} \scriptsize 6};
            
            \draw[white, fill=white] (1.6003e+3,2e-4*3) circle (1.6mm);
            \draw[draw opacity=0.9, fill=white, draw=red_80, line width=.7] (1.6003e+3,2e-4*3) circle (1.4mm) node[] {\color{red_80} \scriptsize \hspace{.3mm}6\hspace{-.2mm}'};

        \end{scope}

        \draw[black, line width=1pt] (current axis.south west) rectangle (current axis.north east);
       
    \end{loglogaxis}
\end{tikzpicture}   
    }
    \\[-0.32cm]
    \subfloat
    {
        \begin{tikzpicture}
   
    \clip(-1.52cm,-.8cm) rectangle (\subfigWidth-1.52cm,\subfigHeight-.8cm);
    \begin{loglogaxis}[%
        title={$N\!=\!8$ antennas --- diagonal direction},
        title style={at={(axis description cs:.5,.95)},anchor=south},
        width=6.75cm,
        height=3.45cm,
        scale only axis,
        xmin=1e-1,
        xmax=1e4,
        ymin=1e-4,
        ymax=1,
        xlabel={$r/\lambda$},
        x label style={at={(axis description cs:.5,-.09)},anchor=north},
        ylabel ={$\epsilon_{90^\circ\!,45^\circ}(r)$},
        grid=minor,
        tick style={line width=.5pt},
        ytick = {1, 1e-1, 1e-2, 1e-3, 1e-4, 1e-6},
        clip=false,
        minor ytick = {1e-6, 2e-6, 3e-6, 4e-6, 5e-6, 6e-6, 7e-6, 8e-6, 9e-6, 1e-5, 2e-5, 3e-5, 4e-5, 5e-5, 6e-5, 7e-5, 8e-5, 9e-5, 1e-4, 2e-4, 3e-4, 4e-4, 5e-4, 6e-4, 7e-4, 8e-4, 9e-4, 1e-3, 2e-3, 3e-3, 4e-3, 5e-3, 6e-3, 7e-3, 8e-3, 9e-3, 1e-2, 2e-2, 3e-2, 4e-2, 5e-2, 6e-2, 7e-2, 8e-2, 9e-2, 1e-1, 2e-1, 3e-1, 4e-1, 5e-1, 6e-1, 7e-1, 8e-1, 9e-1, 1, 2, 3, 4, 5},
        every outer y axis line/.append style={black,very thick},
        legend style={at={(0.97,0.97)},anchor=north east,legend cell align=left, align=left, draw=black,thick}
        ]

        \begin{scope}
            \clip (current axis.south west) rectangle (current axis.north east);
        
            \addplot[black!70, line width=1pt, dash pattern=on 7pt off 1pt on 1pt off 1pt] table [col sep=comma] {tikzfigs/data/d_inf_8_45_ff.dat};

            \addplot[black!70, line width=1pt, dash pattern=on 5pt off 1pt on 1pt off 1pt on 1pt off 1pt] table [col sep=comma] {tikzfigs/data/d_inf_8_45_nf.dat};
            
            \addplot[black!70, line width=1pt, dash pattern=on 6pt off 4pt] table [col sep=comma] {tikzfigs/data/d_2_8_45.dat};
            
            \addplot[black!70, line width=1pt, dash pattern=on 1pt off 1pt] table [col sep=comma] {tikzfigs/data/p_8_45.dat};

            \draw[opacity=0.9, blue, line width=1] (24.5000,1) -- (24.5000,1e-6);

            \draw[opacity=0.9, petrol, line width=1] (13.5560,1) -- (13.5560,1e-6);

            \draw[opacity=0.9, green_120, line width=1] (71.2612,1) -- (71.2612,1e-6);
           
            \draw[opacity=0.9, green_80, line width=1] (33.3060,1) -- (33.3060,1e-6);

            \draw[opacity=0.9, bronze_120, line width=1] (12.9453,1) -- (12.9453,1e-6);

            \draw[opacity=0.9, bronze_80, line width=1] (6.3358,1) -- (6.3358,1e-6);
          
            \draw[opacity=0.9, purple_120, line width=1] (10000,1) -- (10000,1e-6);
           
            \draw[opacity=0.9, purple_80, line width=1] (1.3065,1) -- (1.3065,1e-6);

            \draw[opacity=0.9, red_120, line width=1] (560.7235,1) -- (560.7235,1e-6);
            
            \draw[opacity=0.9, red_80, line width=1] (177.1121,1) -- (177.1121,1e-6);

            \draw[white, fill=white] (24.5000,2e-3*3*3) circle (1.6mm);
            \draw[draw opacity=0.9, fill=white, draw=blue, line width=.7] (24.5000,2e-3*3*3) circle (1.4mm) node[] {\color{blue} \scriptsize 1};
            
            \draw[white, fill=white] (13.5560,7e-3/3) circle (1.6mm);
            \draw[draw opacity=0.9, fill=white, draw=petrol, line width=.7] (13.5560,7e-3/3) circle (1.4mm) node[] {\color{petrol} \scriptsize 2};
            
            \draw[white, fill=white] (71.2612,4e-4*3) circle (1.6mm);
            \draw[draw opacity=0.9, fill=white, draw=green_120, line width=.7] (71.2612,4e-4*3) circle (1.4mm) node[] {\color{green_120} \scriptsize 3};

            \draw[white, fill=white] (33.3060,2e-3*3) circle (1.6mm);
            \draw[draw opacity=0.9, fill=white, draw=green_80, line width=.7] (33.3060,2e-3*3) circle (1.4mm) node[] {\color{green_80} \scriptsize \hspace{.3mm}3\hspace{-.2mm}'};

            \draw[white, fill=white] (12.9453,7e-3/3/3) circle (1.6mm);
            \draw[draw opacity=0.9, fill=white, draw=bronze_120, line width=.7] (12.9453,7e-3/3/3) circle (1.4mm) node[] {\color{bronze_120} \scriptsize 4};

            \draw[white, fill=white] (6.3358,3e-2/3) circle (1.6mm);
            \draw[draw opacity=0.9, fill=white, draw=bronze_80, line width=.7] (6.3358,3e-2/3) circle (1.4mm) node[] {\color{bronze_80} \scriptsize \hspace{.3mm}4\hspace{-.2mm}'};
 
            \draw [purple_120, line width=.8, arrows = {-Stealth[inset=0, length=4pt, angle'=65]}] (4000,2e-4) -- (4000,2e-4*3) -- (9000,2e-4*3);
            \draw[white, fill=white] (4000,2e-4) circle (1.6mm);
            \draw[draw opacity=0.9, fill=white, draw=purple_120, line width=.7] (4000,2e-4) circle (1.4mm) node[] {\color{purple_120} \scriptsize 5};
    
            \draw[white, fill=white] (1.3065,0.2/2) circle (1.6mm);
            \draw[draw opacity=0.9, fill=white, draw=purple_80, line width=.7] (1.3065,0.2/2) circle (1.4mm) node[] {\color{purple_80} \scriptsize \hspace{.3mm}5\hspace{-.2mm}'};
    
            \draw[white, fill=white] (560.7235,2e-4) circle (1.6mm);
            \draw[draw opacity=0.9, fill=white, draw=red_120, line width=.7] (560.7235,2e-4) circle (1.4mm) node[] {\color{red_120} \scriptsize 6};
            
            \draw[white, fill=white] (177.1121,2e-4) circle (1.6mm);
            \draw[draw opacity=0.9, fill=white, draw=red_80, line width=.7] (177.1121,2e-4) circle (1.4mm) node[] {\color{red_80} \scriptsize \hspace{.3mm}6\hspace{-.2mm}'};

        \end{scope}

        \draw[black, line width=1pt] (current axis.south west) rectangle (current axis.north east);
       
    \end{loglogaxis}
\end{tikzpicture}   
    }
    \hspace{-4mm}
    \subfloat
    {
        \begin{tikzpicture}
   
    \clip(-1.52cm,-.8cm) rectangle (\subfigWidth-1.52cm,\subfigHeight-.8cm);
    \begin{loglogaxis}[%
        title={$N\!=\!64$ antennas --- diagonal direction},
        title style={at={(axis description cs:.5,.95)},anchor=south},
        width=6.75cm,
        height=3.45cm,
        scale only axis,
        xmin=1e-1,
        xmax=1e4,
        ymin=1e-4,
        ymax=1,
        xlabel={$r/\lambda$},
        x label style={at={(axis description cs:.5,-.09)},anchor=north},
        ylabel ={$\epsilon_{90^\circ\!,45^\circ}(r)$},
        grid=minor,
        tick style={line width=.5pt},
        ytick = {1, 1e-1, 1e-2, 1e-3, 1e-4, 1e-6},
        clip=false,
        minor ytick = {1e-6, 2e-6, 3e-6, 4e-6, 5e-6, 6e-6, 7e-6, 8e-6, 9e-6, 1e-5, 2e-5, 3e-5, 4e-5, 5e-5, 6e-5, 7e-5, 8e-5, 9e-5, 1e-4, 2e-4, 3e-4, 4e-4, 5e-4, 6e-4, 7e-4, 8e-4, 9e-4, 1e-3, 2e-3, 3e-3, 4e-3, 5e-3, 6e-3, 7e-3, 8e-3, 9e-3, 1e-2, 2e-2, 3e-2, 4e-2, 5e-2, 6e-2, 7e-2, 8e-2, 9e-2, 1e-1, 2e-1, 3e-1, 4e-1, 5e-1, 6e-1, 7e-1, 8e-1, 9e-1, 1, 2, 3, 4, 5},
        every outer y axis line/.append style={black,very thick},
        legend style={at={(0.97,0.97)},anchor=north east,legend cell align=left, align=left, draw=black,thick}
        ]

        \begin{scope}
            \clip (current axis.south west) rectangle (current axis.north east);
        
            \addplot[black!70, line width=1pt, dash pattern=on 7pt off 1pt on 1pt off 1pt] table [col sep=comma] {tikzfigs/data/d_inf_64_45_ff.dat};

            \addplot[black!70, line width=1pt, dash pattern=on 5pt off 1pt on 1pt off 1pt on 1pt off 1pt] table [col sep=comma] {tikzfigs/data/d_inf_64_45_nf.dat};
            
            \addplot[black!70, line width=1pt, dash pattern=on 6pt off 4pt] table [col sep=comma] {tikzfigs/data/d_2_64_45.dat};
            
            \addplot[black!70, line width=1pt, dash pattern=on 1pt off 1pt] table [col sep=comma] {tikzfigs/data/p_64_45.dat};

            \draw[opacity=0.9, blue, line width=1] (1.9845e+03,1) -- (1.9845e+03,1e-6);

            \draw[opacity=0.9, petrol, line width=1] (1.0093e+03,1) -- (1.0093e+03,1e-6);

            \draw[opacity=0.9, green_120, line width=1] (636.5079,1) -- (636.5079,1e-6);
           
            \draw[opacity=0.9, green_80, line width=1] (300.9390,1) -- (300.9390,1e-6);

            \draw[opacity=0.9, bronze_120, line width=1] (849.0415,1) -- (849.0415,1e-6);

            \draw[opacity=0.9, bronze_80, line width=1] (383.3395,1) -- (383.3395,1e-6);
          
           
            \draw[opacity=0.9, purple_80, line width=1] (10.5202,1) -- (10.5202,1e-6);

            \draw[opacity=0.9, red_120, line width=1] (5.0665e+3,1e-4) -- (5.0665e+3,1.5e-1);
            \draw[opacity=0.9, red_120, line width=1] (5.0665e+3,1.85e-1) -- (5.0665e+3,1);

            \draw[opacity=0.9, red_80, line width=1] (1.6003e+03,1) -- (1.6003e+03,1e-6);

            \draw[white, fill=white] (1.9845e+03,1e-3*3) circle (1.6mm);
            \draw[draw opacity=0.9, fill=white, draw=blue, line width=.7] (1.9845e+03,1e-3*3) circle (1.4mm) node[] {\color{blue} \scriptsize 1};
            
            \draw[white, fill=white] (1.0093e+03,4e-3/3) circle (1.6mm);
            \draw[draw opacity=0.9, fill=white, draw=petrol, line width=.7] (1.0093e+03,4e-3/3) circle (1.4mm) node[] {\color{petrol} \scriptsize 2};
            
            \draw[white, fill=white] (636.5079,1e-2/3) circle (1.6mm);
            \draw[draw opacity=0.9, fill=white, draw=green_120, line width=.7] (636.5079,1e-2/3) circle (1.4mm) node[] {\color{green_120} \scriptsize 3};

            \draw[white, fill=white] (300.9390,4e-2/3) circle (1.6mm);
            \draw[draw opacity=0.9, fill=white, draw=green_80, line width=.7] (300.9390,4e-2/3) circle (1.4mm) node[] {\color{green_80} \scriptsize \hspace{.3mm}3\hspace{-.2mm}'};

            \draw[white, fill=white] (849.0415,4e-3/3/3) circle (1.6mm);
            \draw[draw opacity=0.9, fill=white, draw=bronze_120, line width=.7] (849.0415,4e-3/3/3) circle (1.4mm) node[] {\color{bronze_120} \scriptsize 4};

            \draw[white, fill=white] (383.3395,3e-2/6) circle (1.6mm);
            \draw[draw opacity=0.9, fill=white, draw=bronze_80, line width=.7] (383.3395,3e-2/6) circle (1.4mm) node[] {\color{bronze_80} \scriptsize \hspace{.3mm}4\hspace{-.2mm}'};
 
            \draw [purple_120, line width=.8, arrows = {-Stealth[inset=0, length=4pt, angle'=65]}] (4000,5e-1) -- (4000,5e-1/3) -- (9000,5e-1/3);
            \draw[white, fill=white] (4000,5e-1) circle (1.6mm);
            \draw[draw opacity=0.9, fill=white, draw=purple_120, line width=.7] (4000,5e-1) circle (1.4mm) node[] {\color{purple_120} \scriptsize 5};
    
            \draw[white, fill=white] ( 10.5202,8e-1/3) circle (1.6mm);
            \draw[draw opacity=0.9, fill=white, draw=purple_80, line width=.7] ( 10.5202,8e-1/3) circle (1.4mm) node[] {\color{purple_80} \scriptsize \hspace{.3mm}5\hspace{-.2mm}'};
    
            \draw[white, fill=white] (5.0665e+03,2e-4*3) circle (1.6mm);
            \draw[draw opacity=0.9, fill=white, draw=red_120, line width=.7] (5.0665e+03,2e-4*3) circle (1.4mm) node[] {\color{red_120} \scriptsize 6};
            
            \draw[white, fill=white] (1.6003e+03,4e-3/3/3) circle (1.6mm);
            \draw[draw opacity=0.9, fill=white, draw=red_80, line width=.7] (1.6003e+03,4e-3/3/3) circle (1.4mm) node[] {\color{red_80} \scriptsize \hspace{.3mm}6\hspace{-.2mm}'};

        \end{scope}

        \draw[black, line width=1pt] (current axis.south west) rectangle (current axis.north east);
       
    \end{loglogaxis}
\end{tikzpicture}   
    }
    \\[-0.32cm]
    \subfloat
    {
        \begin{tikzpicture}
   
    \clip(-1.52cm,-.8cm) rectangle (\subfigWidth-1.52cm,\subfigHeight-.8cm);
    \begin{loglogaxis}[%
        title={$N\!=\!8$ antennas --- side direction},
        title style={at={(axis description cs:.5,.95)},anchor=south},
        width=6.75cm,
        height=3.45cm,
        scale only axis,
        xmin=1e-1,
        xmax=1e4,
        ymin=1e-4,
        ymax=1,
        xlabel={$r/\lambda$},
        x label style={at={(axis description cs:.5,-.09)},anchor=north},
        ylabel ={$\epsilon_{90^\circ\!,90^\circ}(r)$},
        grid=minor,
        tick style={line width=.5pt},
        ytick = {1, 1e-1, 1e-2, 1e-3, 1e-4, 1e-6},
        clip=false,
        minor ytick = {1e-6, 2e-6, 3e-6, 4e-6, 5e-6, 6e-6, 7e-6, 8e-6, 9e-6, 1e-5, 2e-5, 3e-5, 4e-5, 5e-5, 6e-5, 7e-5, 8e-5, 9e-5, 1e-4, 2e-4, 3e-4, 4e-4, 5e-4, 6e-4, 7e-4, 8e-4, 9e-4, 1e-3, 2e-3, 3e-3, 4e-3, 5e-3, 6e-3, 7e-3, 8e-3, 9e-3, 1e-2, 2e-2, 3e-2, 4e-2, 5e-2, 6e-2, 7e-2, 8e-2, 9e-2, 1e-1, 2e-1, 3e-1, 4e-1, 5e-1, 6e-1, 7e-1, 8e-1, 9e-1, 1, 2, 3, 4, 5},
        every outer y axis line/.append style={black,very thick},
        legend style={at={(0.97,0.97)},anchor=north east,legend cell align=left, align=left, draw=black,thick}
        ]

        \begin{scope}
            \clip (current axis.south west) rectangle (current axis.north east);
        
            \addplot[black!70, line width=1pt, dash pattern=on 7pt off 1pt on 1pt off 1pt] table [col sep=comma] {tikzfigs/data/d_inf_8_90_ff.dat};

            \addplot[black!70, line width=1pt, dash pattern=on 5pt off 1pt on 1pt off 1pt on 1pt off 1pt] table [col sep=comma] {tikzfigs/data/d_inf_8_90_nf.dat};
            
            \addplot[black!70, line width=1pt, dash pattern=on 6pt off 4pt] table [col sep=comma] {tikzfigs/data/d_2_8_90.dat};

            \addplot[black!70, line width=1pt, dash pattern=on 1pt off 1pt] table [col sep=comma] {tikzfigs/data/p_8_90.dat};

            \draw[opacity=0.9, blue, line width=1] (24.5000,1) -- (24.5000,1e-6);

            \draw[opacity=0.9, petrol, line width=1] (1.7228,1) -- (1.7228,1e-6);

            \draw[opacity=0.9, green_120, line width=1] (100.6939,1) -- (100.6939,1e-6);
           
            \draw[opacity=0.9, green_80, line width=1] ( 47.6078,1) -- ( 47.6078,1e-6);

            \draw[opacity=0.9, bronze_120, line width=1] ( 1.7031,1) -- ( 1.7031,1e-6);

            \draw[opacity=0.9, bronze_80, line width=1] (1.6836,1) -- (1.6836,1e-6);
          
            \draw[opacity=0.9, purple_120, line width=1] (100000,1) -- (100000,1e-6);
           
            \draw[opacity=0.9, purple_80, line width=1] (0.1821,1) -- (0.1821,1e-6);

            \draw[opacity=0.9, red_120, line width=1] ( 560.7235,1) -- ( 560.7235,1e-6);
            
            \draw[opacity=0.9, red_80, line width=1] ( 177.1121,1) -- ( 177.1121,1e-6);

            \draw[white, fill=white] (24.5000,2e-4*3) circle (1.6mm);
            \draw[draw opacity=0.9, fill=white, draw=blue, line width=.7] (24.5000,2e-4*3) circle (1.4mm) node[] {\color{blue} \scriptsize 1};
            
            \draw[white, fill=white] (1.7228,1e-1/3) circle (1.6mm);
            \draw[draw opacity=0.9, fill=white, draw=petrol, line width=.7] (1.7228,1e-1/3) circle (1.4mm) node[] {\color{petrol} \scriptsize 2};
            
            \draw[white, fill=white] (100.6939,2e-4) circle (1.6mm);
            \draw[draw opacity=0.9, fill=white, draw=green_120, line width=.7] (100.6939,2e-4) circle (1.4mm) node[] {\color{green_120} \scriptsize 3};

            \draw[white, fill=white] ( 47.6078,2e-4) circle (1.6mm);
            \draw[draw opacity=0.9, fill=white, draw=green_80, line width=.7] ( 47.6078,2e-4) circle (1.4mm) node[] {\color{green_80} \scriptsize \hspace{.3mm}3\hspace{-.2mm}'};

            \draw[white, fill=white] ( 1.7031,1e-1/3/3) circle (1.6mm);
            \draw[draw opacity=0.9, fill=white, draw=bronze_120, line width=.7] ( 1.7031,1e-1/3/3) circle (1.4mm) node[] {\color{bronze_120} \scriptsize 4};

            \draw[white, fill=white] (1.6836,1e-1/3/3/3) circle (1.6mm);
            \draw[draw opacity=0.9, fill=white, draw=bronze_80, line width=.7] (1.6836,1e-1/3/3/3) circle (1.4mm) node[] {\color{bronze_80} \scriptsize \hspace{.3mm}4\hspace{-.2mm}'};
 
            \draw [purple_120, line width=.8, arrows = {-Stealth[inset=0, length=4pt, angle'=65]}] (4000,2e-4) -- (4000,2e-4*3) -- (9000,2e-4*3);
            \draw[white, fill=white] (4000,2e-4) circle (1.6mm);
            \draw[draw opacity=0.9, fill=white, draw=purple_120, line width=.7] (4000,2e-4) circle (1.4mm) node[] {\color{purple_120} \scriptsize 5};
    
            \draw[white, fill=white] ( 0.1821,3e-1/3) circle (1.6mm);
            \draw[draw opacity=0.9, fill=white, draw=purple_80, line width=.7] ( 0.1821,3e-1/3) circle (1.4mm) node[] {\color{purple_80} \scriptsize \hspace{.3mm}5\hspace{-.2mm}'};
    
            \draw[white, fill=white] ( 560.7235,2e-4) circle (1.6mm);
            \draw[draw opacity=0.9, fill=white, draw=red_120, line width=.7] ( 560.7235,2e-4) circle (1.4mm) node[] {\color{red_120} \scriptsize 6};
            
            \draw[white, fill=white] ( 177.1121,2e-4) circle (1.6mm);
            \draw[draw opacity=0.9, fill=white, draw=red_80, line width=.7] ( 177.1121,2e-4) circle (1.4mm) node[] {\color{red_80} \scriptsize \hspace{.3mm}6\hspace{-.2mm}'};

        \end{scope}

        \draw[black, line width=1pt] (current axis.south west) rectangle (current axis.north east);
       
    \end{loglogaxis}
\end{tikzpicture}   
    }
    \hspace{-4mm}
    \subfloat
    {
        \begin{tikzpicture}
   
    \clip(-1.52cm,-.8cm) rectangle (\subfigWidth-1.52cm,\subfigHeight-.8cm);
    \begin{loglogaxis}[%
        title={$N\!=\!64$ antennas --- side direction},
        title style={at={(axis description cs:.5,.95)},anchor=south},
        width=6.75cm,
        height=3.45cm,
        scale only axis,
        xmin=1e-1,
        xmax=1e4,
        ymin=1e-4,
        ymax=1,
        xlabel={$r/\lambda$},
        x label style={at={(axis description cs:.5,-.09)},anchor=north},
        ylabel ={$\epsilon_{90^\circ\!,90^\circ}(r)$},
        grid=minor,
        tick style={line width=.5pt},
        ytick = {1, 1e-1, 1e-2, 1e-3, 1e-4, 1e-6},
        clip=false,
        minor ytick = {1e-6, 2e-6, 3e-6, 4e-6, 5e-6, 6e-6, 7e-6, 8e-6, 9e-6, 1e-5, 2e-5, 3e-5, 4e-5, 5e-5, 6e-5, 7e-5, 8e-5, 9e-5, 1e-4, 2e-4, 3e-4, 4e-4, 5e-4, 6e-4, 7e-4, 8e-4, 9e-4, 1e-3, 2e-3, 3e-3, 4e-3, 5e-3, 6e-3, 7e-3, 8e-3, 9e-3, 1e-2, 2e-2, 3e-2, 4e-2, 5e-2, 6e-2, 7e-2, 8e-2, 9e-2, 1e-1, 2e-1, 3e-1, 4e-1, 5e-1, 6e-1, 7e-1, 8e-1, 9e-1, 1, 2, 3, 4, 5},
        every outer y axis line/.append style={black,very thick},
        legend style={at={(0.97,0.97)},anchor=north east,legend cell align=left, align=left, draw=black,thick}
        ]

        \begin{scope}
            \clip (current axis.south west) rectangle (current axis.north east);
        
            \addplot[black!70, line width=1pt, dash pattern=on 7pt off 1pt on 1pt off 1pt] table [col sep=comma] {tikzfigs/data/d_inf_64_90_ff.dat};

            \addplot[black!70, line width=1pt, dash pattern=on 5pt off 1pt on 1pt off 1pt on 1pt off 1pt] table [col sep=comma] {tikzfigs/data/d_inf_64_90_nf.dat};
            
            \addplot[black!70, line width=1pt, dash pattern=on 6pt off 4pt] table [col sep=comma] {tikzfigs/data/d_2_64_90.dat};
            
            \addplot[black!70, line width=1pt, dash pattern=on 1pt off 1pt] table [col sep=comma] {tikzfigs/data/p_64_90.dat};

            \draw[opacity=0.9, blue, line width=1] (1.9845e+03,1) -- (1.9845e+03,1e-6);

            \draw[opacity=0.9, petrol, line width=1] (15.7470,1) -- (15.7470,1e-6);

            \draw[opacity=0.9, green_120, line width=1] (899.4022,1) -- (899.4022,1e-6);
           
            \draw[opacity=0.9, green_80, line width=1] (425.2346,1) -- (425.2346,1e-6);

            \draw[opacity=0.9, bronze_120, line width=1] (15.5665,1) -- (15.5665,1e-6);

            \draw[opacity=0.9, bronze_80, line width=1] (15.5665,1) -- (15.5665,1e-6);
          
           
            \draw[opacity=0.9, purple_80, line width=1] (1.0865,1) -- (1.0865,1e-6);

            \draw[opacity=0.9, red_120, line width=1] (5.0665e+3,1e-4) -- (5.0665e+3,1.5e-1);
            \draw[opacity=0.9, red_120, line width=1] (5.0665e+3,1.85e-1) -- (5.0665e+3,1);
            
            \draw[opacity=0.9, red_80, line width=1] (1.6003e+03,1) -- (1.6003e+03,1e-6);

            \draw[white, fill=white] (1.9845e+03,2e-4*3) circle (1.6mm);
            \draw[draw opacity=0.9, fill=white, draw=blue, line width=.7] (1.9845e+03,2e-4*3) circle (1.4mm) node[] {\color{blue} \scriptsize 1};
            
            \draw[white, fill=white] (15.7470,1e-2) circle (1.6mm);
            \draw[draw opacity=0.9, fill=white, draw=petrol, line width=.7] (15.7470,1e-2) circle (1.4mm) node[] {\color{petrol} \scriptsize 2};
            
            \draw[white, fill=white] (899.4022,2e-4) circle (1.6mm);
            \draw[draw opacity=0.9, fill=white, draw=green_120, line width=.7] (899.4022,2e-4) circle (1.4mm) node[] {\color{green_120} \scriptsize 3};

            \draw[white, fill=white] (425.2346,2e-4) circle (1.6mm);
            \draw[draw opacity=0.9, fill=white, draw=green_80, line width=.7] (425.2346,2e-4) circle (1.4mm) node[] {\color{green_80} \scriptsize \hspace{.3mm}3\hspace{-.2mm}'};

            \draw[white, fill=white] (15.5665,1e-2/3) circle (1.6mm);
            \draw[draw opacity=0.9, fill=white, draw=bronze_120, line width=.7] (15.5665,1e-2/3) circle (1.4mm) node[] {\color{bronze_120} \scriptsize 4};

            \draw[white, fill=white] (15.5665,1e-2/9) circle (1.6mm);
            \draw[draw opacity=0.9, fill=white, draw=bronze_80, line width=.7] (15.5665,1e-2/9) circle (1.4mm) node[] {\color{bronze_80} \scriptsize \hspace{.3mm}4\hspace{-.2mm}'};
 
            \draw [purple_120, line width=.8, arrows = {-Stealth[inset=0, length=4pt, angle'=65]}] (4000,5e-1) -- (4000,5e-1/3) -- (9000,5e-1/3);
            \draw[white, fill=white] (4000,5e-1) circle (1.6mm);
            \draw[draw opacity=0.9, fill=white, draw=purple_120, line width=.7] (4000,5e-1) circle (1.4mm) node[] {\color{purple_120} \scriptsize 5};
    
            \draw[white, fill=white] (1.0865,1e-1) circle (1.6mm);
            \draw[draw opacity=0.9, fill=white, draw=purple_80, line width=.7] (1.0865,1e-1) circle (1.4mm) node[] {\color{purple_80} \scriptsize \hspace{.3mm}5\hspace{-.2mm}'};
    
            \draw[white, fill=white] (5.0665e+03,2e-4) circle (1.6mm);
            \draw[draw opacity=0.9, fill=white, draw=red_120, line width=.7] (5.0665e+03,2e-4) circle (1.4mm) node[] {\color{red_120} \scriptsize 6};
            
            \draw[white, fill=white] (1.6003e+03,2e-4) circle (1.6mm);
            \draw[draw opacity=0.9, fill=white, draw=red_80, line width=.7] (1.6003e+03,2e-4) circle (1.4mm) node[] {\color{red_80} \scriptsize \hspace{.3mm}6\hspace{-.2mm}'};

        \end{scope}

        \draw[black, line width=1pt] (current axis.south west) rectangle (current axis.north east);
       
    \end{loglogaxis}
\end{tikzpicture}   
    }
    \caption{The framework introduced in~\fref{sec:framework}, when applied to the test scenarios described in~\fref{sec:test_scenarios}, evaluates the near–far-field boundaries listed in~\fref{tab:distances_under_test}. In all antenna array scenarios, the element spacing was set to $d=\lambda/2$. The curves for infinitesimal dipole sources are computed from closed-form expressions; those for half-wave dipole and patch antennas are derived from fields obtained via full-wave electromagnetic simulation. None of the tested near-far-field boundaries is able to accurately and consistently predict the position at which the approximation error convergence to zero.}\label{fig:results_I}
\end{figure*}
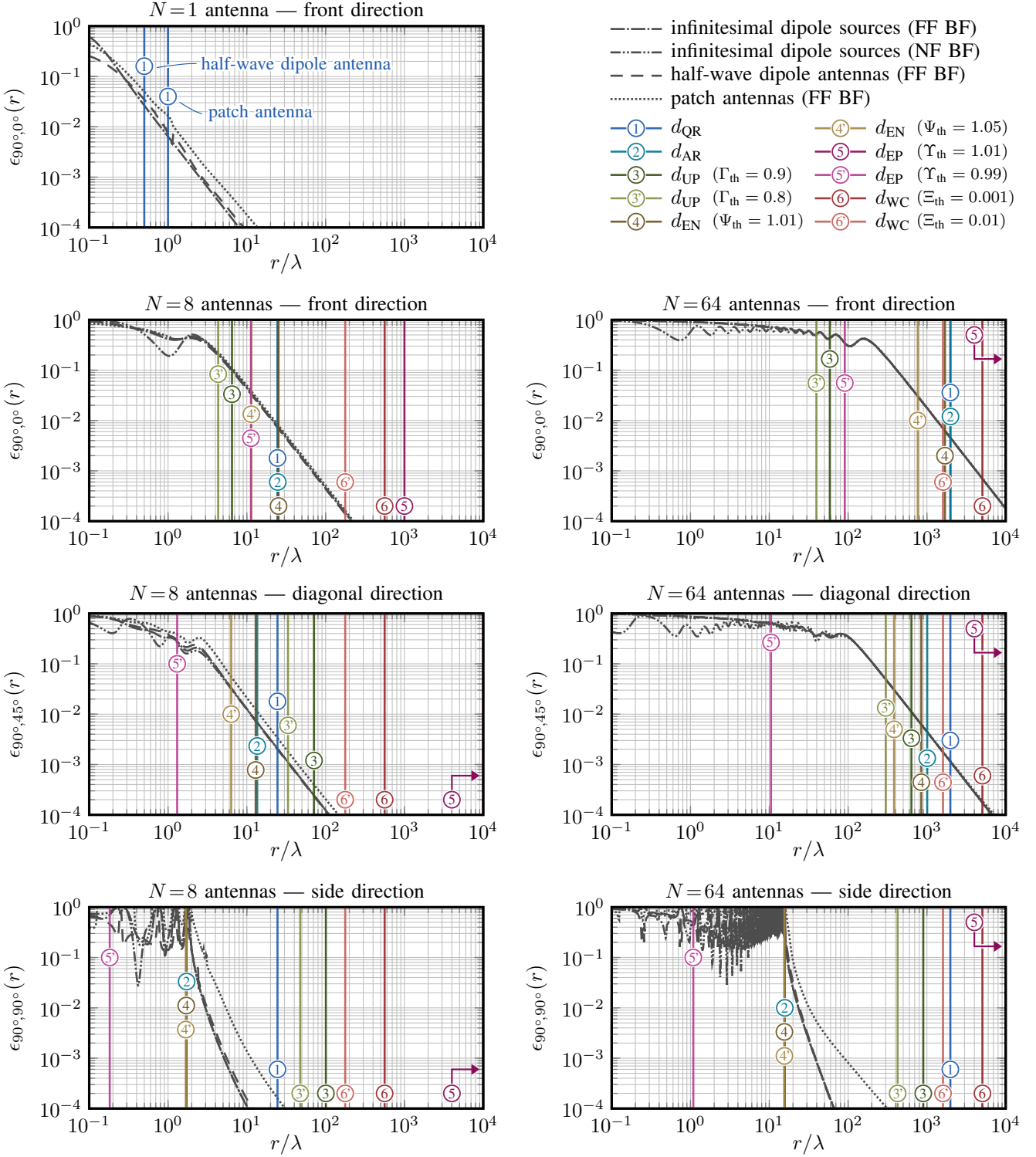
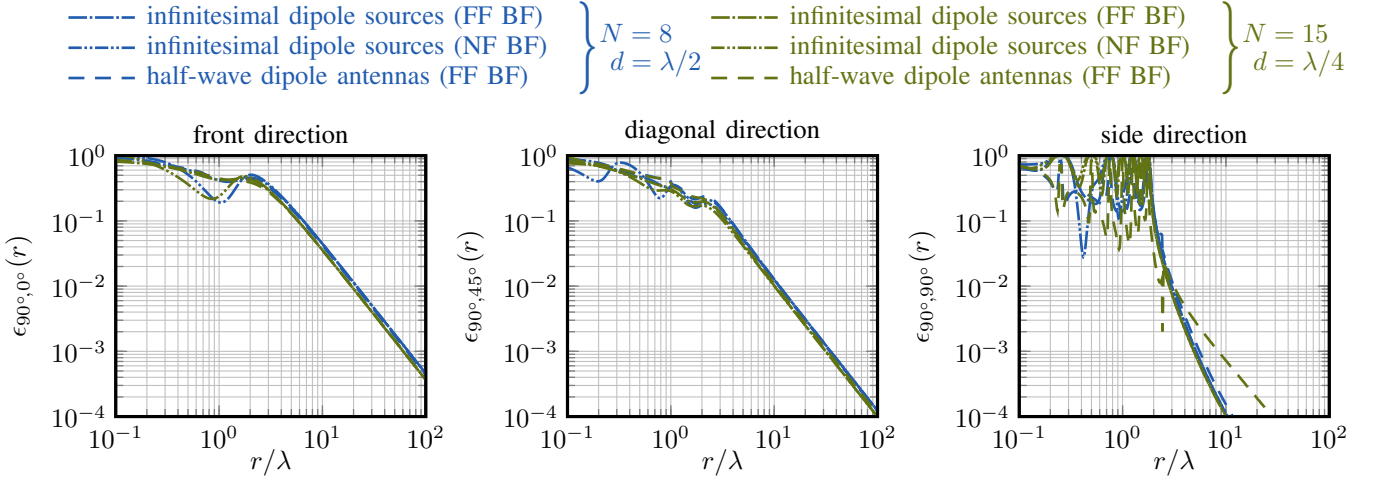
\begin{figure*}[htp!]
    \subfloat
    {
        \begin{tikzpicture}
            \clip(-1.55,2.1) rectangle (16.5,4);
            \begin{scope}[shift={(-6mm,0)}]

                \begin{scope}[shift={(6.9cm,3.1cm)}];   
                    \node[anchor=west] at (0pt,0) {\color{blue}$N=8$};
                    \node[anchor=west] at (4pt,-4mm) {\color{blue}$d=\lambda/2$};
                \end{scope}

                \begin{scope}[shift={(6.6cm,2.9cm)}];   
                    \node[anchor=west] at (0pt,0) {\color{blue}$\scalebox{1.6}[3.5]{\}}$};
                \end{scope}
            
                \begin{scope}[shift={(0cm,3.3cm)}];   
                    \draw[blue, line width=1pt, dash pattern=on 7pt off 1pt on 1pt off 1pt] (0,0) -- (25pt,0);
                    \node[anchor=west] at (26pt,0) {\color{blue} infinitesimal dipole sources (FF BF)};
                \end{scope}
        
                \begin{scope}[shift={(0cm,2.9cm)}];   
                    \draw[blue, line width=1pt, dash pattern=on 5pt off 1pt on 1pt off 1pt on 1pt off 1pt] (0,0) -- (25pt,0);
                    \node[anchor=west] at (26pt,0) {\color{blue} infinitesimal dipole sources (NF BF)};
                \end{scope}
        
                \begin{scope}[shift={(0cm,2.5cm)}];   
                    \draw[blue, line width=1pt, dash pattern=on 6pt off 4pt] (0,0) -- (25pt,0);
                    \node[anchor=west] at (26pt,0) {\color{blue} half-wave dipole antennas (FF BF)};
                \end{scope}

                \begin{scope}[shift={(15.4cm,3.1cm)}];   
                    \node[anchor=west] at (0pt,0) {\color{green}$N=15$};
                    \node[anchor=west] at (4pt,-4mm) {\color{green}$d=\lambda/4$};
                \end{scope}

                \begin{scope}[shift={(15.1cm,2.9cm)}];   
                    \node[anchor=west] at (0pt,0) {\color{green}$\scalebox{1.6}[3.5]{\}}$};
                \end{scope}
            
                \begin{scope}[shift={(8.5cm,3.3cm)}];   
                    \draw[green, line width=1pt, dash pattern=on 7pt off 1pt on 1pt off 1pt] (0,0) -- (25pt,0);
                    \node[anchor=west] at (26pt,0) {\color{green} infinitesimal dipole sources (FF BF)};
                \end{scope}
        
                \begin{scope}[shift={(8.5cm,2.9cm)}];   
                    \draw[green, line width=1pt, dash pattern=on 5pt off 1pt on 1pt off 1pt on 1pt off 1pt] (0,0) -- (25pt,0);
                    \node[anchor=west] at (26pt,0) {\color{green} infinitesimal dipole sources (NF BF)};
                \end{scope}
        
                \begin{scope}[shift={(8.5cm,2.5cm)}];   
                    \draw[green, line width=1pt, dash pattern=on 6pt off 4pt] (0,0) -- (25pt,0);
                    \node[anchor=west] at (26pt,0) {\color{green} half-wave dipole antennas (FF BF)};
                \end{scope}
            \end{scope}
        \end{tikzpicture}  
    }
    \\[-0.32cm]
    \subfloat
    {
        \begin{tikzpicture}
           
            \begin{loglogaxis}[%
                title={front direction},
                title style={at={(axis description cs:.5,.95)},anchor=south},
                width=4.1cm,
                height=3.45cm,
                scale only axis,
                xmin=1e-1,
                xmax=1e2,
                ymin=1e-4,
                ymax=1,
                xlabel={$r/\lambda$},
                x label style={at={(axis description cs:.5,-.09)},anchor=north},
                ylabel ={$\epsilon_{90^\circ\!,0^\circ}(r)$},
                grid=minor,
                tick style={line width=.5pt},
                ytick = {1, 1e-1, 1e-2, 1e-3, 1e-4, 1e-5, 1e-6},
                clip=false,
                minor ytick = {1e-6, 2e-6, 3e-6, 4e-6, 5e-6, 6e-6, 7e-6, 8e-6, 9e-6, 1e-5, 2e-5, 3e-5, 4e-5, 5e-5, 6e-5, 7e-5, 8e-5, 9e-5, 1e-4, 2e-4, 3e-4, 4e-4, 5e-4, 6e-4, 7e-4, 8e-4, 9e-4, 1e-3, 2e-3, 3e-3, 4e-3, 5e-3, 6e-3, 7e-3, 8e-3, 9e-3, 1e-2, 2e-2, 3e-2, 4e-2, 5e-2, 6e-2, 7e-2, 8e-2, 9e-2, 1e-1, 2e-1, 3e-1, 4e-1, 5e-1, 6e-1, 7e-1, 8e-1, 9e-1, 1, 2, 3, 4, 5},
                every outer y axis line/.append style={black,very thick},
                legend style={at={(0.97,0.97)},anchor=north east,legend cell align=left, align=left, draw=black,thick}
                ]
        
                \begin{scope}
                    \clip (current axis.south west) rectangle (current axis.north east);

                    \addplot[blue, line width=1pt, dash pattern=on 7pt off 1pt on 1pt off 1pt] table [col sep=comma] {tikzfigs/data/d_inf_8_0_ff.dat};

                    \addplot[blue, line width=1pt, dash pattern=on 7pt off 1pt on 1pt off 1pt on 1pt off 1pt] table [col sep=comma] {tikzfigs/data/d_inf_8_0_nf.dat};
                    
                    \addplot[blue, line width=1pt, dash pattern=on 6pt off 4pt] table [col sep=comma] {tikzfigs/data/d_2_8_0.dat};

                    \addplot[green, line width=1pt, dash pattern=on 7pt off 1pt on 1pt off 1pt] table [col sep=comma] {tikzfigs/data/d_inf_dense_15_0_ff.dat};

                    \addplot[green, line width=1pt, dash pattern=on 7pt off 1pt on 1pt off 1pt on 1pt off 1pt] table [col sep=comma] {tikzfigs/data/d_inf_dense_15_0_nf.dat};
                    
                    \addplot[green, line width=1pt, dash pattern=on 6pt off 4pt] table [col sep=comma] {tikzfigs/data/d_2_dense_15_0.dat};

                \end{scope}
        
                \draw[black, line width=1pt] (current axis.south west) rectangle (current axis.north east);
               
            \end{loglogaxis}
        \end{tikzpicture} 
    }
    \hspace{-4mm}
    \subfloat
    {
        \begin{tikzpicture}
           
            \begin{loglogaxis}[%
                title={diagonal direction},
                title style={at={(axis description cs:.5,.95)},anchor=south},
                width=4.1cm,
                height=3.45cm,
                scale only axis,
                xmin=1e-1,
                xmax=1e2,
                ymin=1e-4,
                ymax=1,
                xlabel={$r/\lambda$},
                x label style={at={(axis description cs:.5,-.09)},anchor=north},
                ylabel ={$\epsilon_{90^\circ\!,45^\circ}(r)$},
                grid=minor,
                tick style={line width=.5pt},
                ytick = {1, 1e-1, 1e-2, 1e-3, 1e-4, 1e-5, 1e-6},
                clip=false,
                minor ytick = {1e-6, 2e-6, 3e-6, 4e-6, 5e-6, 6e-6, 7e-6, 8e-6, 9e-6, 1e-5, 2e-5, 3e-5, 4e-5, 5e-5, 6e-5, 7e-5, 8e-5, 9e-5, 1e-4, 2e-4, 3e-4, 4e-4, 5e-4, 6e-4, 7e-4, 8e-4, 9e-4, 1e-3, 2e-3, 3e-3, 4e-3, 5e-3, 6e-3, 7e-3, 8e-3, 9e-3, 1e-2, 2e-2, 3e-2, 4e-2, 5e-2, 6e-2, 7e-2, 8e-2, 9e-2, 1e-1, 2e-1, 3e-1, 4e-1, 5e-1, 6e-1, 7e-1, 8e-1, 9e-1, 1, 2, 3, 4, 5},
                every outer y axis line/.append style={black,very thick},
                legend style={at={(0.97,0.97)},anchor=north east,legend cell align=left, align=left, draw=black,thick}
                ]
        
                \begin{scope}
                    \clip (current axis.south west) rectangle (current axis.north east);

                    \addplot[blue, line width=1pt, dash pattern=on 7pt off 1pt on 1pt off 1pt] table [col sep=comma] {tikzfigs/data/d_inf_8_45_ff.dat};

                    \addplot[blue, line width=1pt, dash pattern=on 7pt off 1pt on 1pt off 1pt on 1pt off 1pt] table [col sep=comma] {tikzfigs/data/d_inf_8_45_nf.dat};
                    
                    \addplot[blue, line width=1pt, dash pattern=on 6pt off 4pt] table [col sep=comma] {tikzfigs/data/d_2_8_45.dat};

                    \addplot[green, line width=1pt, dash pattern=on 7pt off 1pt on 1pt off 1pt] table [col sep=comma] {tikzfigs/data/d_inf_dense_15_45_ff.dat};

                    \addplot[green, line width=1pt, dash pattern=on 7pt off 1pt on 1pt off 1pt on 1pt off 1pt] table [col sep=comma] {tikzfigs/data/d_inf_dense_15_45_nf.dat};
                    
                    \addplot[green, line width=1pt, dash pattern=on 6pt off 4pt] table [col sep=comma] {tikzfigs/data/d_2_dense_15_45.dat};

                \end{scope}
        
                \draw[black, line width=1pt] (current axis.south west) rectangle (current axis.north east);
               
            \end{loglogaxis}
        \end{tikzpicture} 
    }
    \hspace{-4mm}
    \subfloat
    {
        \begin{tikzpicture}
           
            \begin{loglogaxis}[%
                title={side direction},
                title style={at={(axis description cs:.5,.95)},anchor=south},
                width=4.1cm,
                height=3.45cm,
                scale only axis,
                xmin=1e-1,
                xmax=1e2,
                ymin=1e-4,
                ymax=1,
                xlabel={$r/\lambda$},
                x label style={at={(axis description cs:.5,-.09)},anchor=north},
                ylabel ={$\epsilon_{90^\circ\!,90^\circ}(r)$},
                grid=minor,
                tick style={line width=.5pt},
                ytick = {1, 1e-1, 1e-2, 1e-3, 1e-4, 1e-5, 1e-6},
                clip=false,
                minor ytick = {1e-6, 2e-6, 3e-6, 4e-6, 5e-6, 6e-6, 7e-6, 8e-6, 9e-6, 1e-5, 2e-5, 3e-5, 4e-5, 5e-5, 6e-5, 7e-5, 8e-5, 9e-5, 1e-4, 2e-4, 3e-4, 4e-4, 5e-4, 6e-4, 7e-4, 8e-4, 9e-4, 1e-3, 2e-3, 3e-3, 4e-3, 5e-3, 6e-3, 7e-3, 8e-3, 9e-3, 1e-2, 2e-2, 3e-2, 4e-2, 5e-2, 6e-2, 7e-2, 8e-2, 9e-2, 1e-1, 2e-1, 3e-1, 4e-1, 5e-1, 6e-1, 7e-1, 8e-1, 9e-1, 1, 2, 3, 4, 5},
                every outer y axis line/.append style={black,very thick},
                legend style={at={(0.97,0.97)},anchor=north east,legend cell align=left, align=left, draw=black,thick}
                ]
        
                \begin{scope}
                    \clip (current axis.south west) rectangle (current axis.north east);

                    \addplot[blue, line width=1pt, dash pattern=on 7pt off 1pt on 1pt off 1pt] table [col sep=comma] {tikzfigs/data/d_inf_8_90_ff.dat};

                    \addplot[blue, line width=1pt, dash pattern=on 7pt off 1pt on 1pt off 1pt on 1pt off 1pt] table [col sep=comma] {tikzfigs/data/d_inf_8_90_nf.dat};
                    
                    \addplot[blue, line width=1pt, dash pattern=on 6pt off 4pt] table [col sep=comma] {tikzfigs/data/d_2_8_90.dat};

                    \addplot[green, line width=1pt, dash pattern=on 7pt off 1pt on 1pt off 1pt] table [col sep=comma] {tikzfigs/data/d_inf_dense_15_90_ff.dat};

                    \addplot[green, line width=1pt, dash pattern=on 7pt off 1pt on 1pt off 1pt on 1pt off 1pt] table [col sep=comma] {tikzfigs/data/d_inf_dense_15_90_nf.dat};
                    
                    \addplot[green, line width=1pt, dash pattern=on 6pt off 4pt] table [col sep=comma] {tikzfigs/data/d_2_dense_15_90.dat};

                \end{scope}
        
                \draw[black, line width=1pt] (current axis.south west) rectangle (current axis.north east);
               
            \end{loglogaxis}
        \end{tikzpicture} 
    }
    \caption{Approximation error of uniform linear dipole arrays with a total array length of $Nd = 3.5\lambda$. The results indicate that, in the absence of inter-antenna coupling, the decay behavior of the approximation error is primarily governed by the total dimensions of the antenna system. However, this observation no longer holds when inter-antenna coupling becomes significant, which is the case for the $d=\lambda/4$ separated antenna array.}\label{fig:results_II}
\end{figure*}

We now present the results of the experiments conducted for this paper. 
To this end, we applied the framework introduced in~\fref{sec:framework} to the test scenarios described in~\fref{sec:test_scenarios} and evaluated the near–far-field boundaries listed in~\fref{tab:distances_under_test}. 
For some of the tested near–far-field boundaries, a threshold value must be specified. 
For each of these boundaries, we used at least one threshold value proposed or employed by the respective authors in their original works. 
To assess the influence of the threshold choice, we additionally tested a second value. 
Although we did not systematically optimize the second value thresholds, we selected them in good faith to support robust performance of the boundaries under test.
The quasi-Rayleigh distance depends on the largest dimension of the antenna system $D_\up{S}$. 
In the single antenna scenarios, for the half-wavelength dipole antenna, we used $D_\up{S}=\lambda/2$, and for the patch antenna, we used $D_\up{S}=\lambda/\sqrt{2}$. 
For a single infinitesimal dipole source, it holds that $D_\up{S}=0$, and therefore the quasi-Rayleigh distance is not included for this scenario in our results.
For (uniform linear) arrays with more than one element, we used the approximation $D_\up{S}\approx d N$, where $d$ is the inter-antenna distance and $N$ is the number of antennas. 
The main results are shown in~\fref{fig:results_I}. 
The approximation-error curves of different antenna types exhibit similar behavior when the same excitation scheme is used. 
This similarity increases with larger system size and when the test line is placed in front ($\varphi=0^\circ$) of the array. 
In particular, the approximation-error curves for different antenna types become more alike when moving from $N\!=\!1$ to $N\!=\!8$ and $N\!=\!64$, and from $\varphi\!=\!90^\circ$ to $\varphi\!=\!45^\circ$ and $\varphi\!=\!0^\circ$. 
Not surprisingly, the choice between NF BF and FF BF has a significant influence near the antenna system, but not at larger distances.
Considering the performance of the tested near-far-field boundaries, none of the evaluated near–far-field boundaries is able to accurately and reliably predict the distance at which the approximation error starts to decay. 
Some distances, at least in certain scenarios, lie in regions where the approximation error is still large. 
Following the discussion in~\fref{sec:idea}, this indicates that such boundaries underestimate the distance to the far-field region.  
Other boundaries, by contrast, overestimate the far-field distance by orders of magnitude compared to the point where the approximation-error curve begins to decay. 
In~\fref{fig:results_II}, the approximation error of arrays of $N\!=\!8$ dipoles with an element spacing of $d\!=\!\lambda/2$ is compared to that of arrays of $N=15$ dipoles with a spacing of $d=\lambda/4$, 
i.e., the two array configurations share the same total length. 
In the forward direction, the approximation error curves of the two array configurations are nearly identical. 
This similarity persists in the diagonal direction, although it becomes less pronounced. 
In the side direction, in the near-field region, the curves deviate noticeably, while in the far-field region the curves of the configurations with negligible inter-antenna coupling remain close to each other. 
In contrast, the configuration with strong inter-antenna coupling (half-wave dipoles separated by $\lambda/4$) exhibits a distinctly different decay. 
\begin{rem}
    We highlight that, even though the different antenna types included in our experiments generally produce very similar approximation-error curves, this does not imply that the electromagnetic fields emitted by those antennas are also similar. 
    Indeed, the fields radiated by the highly directive patch antennas differ substantially from those emitted by the rather omnidirectional dipole antennas. 
\end{rem}

\section{Conclusions}

In this paper, we have addressed the question of how to determine the location of the near–far-field transition region for a given antenna system.
To this end, we have proposed a framework based on Maxwell’s equations that can be applied to calculated, simulated, and measured electromagnetic fields.
With our framework one can determine the location of the near–far-field transition region of an arbitrary antenna system in accordance with the field-region definitions provided in the IEEE Standard for Definitions of Terms for Antennas~\cite[Sec.~4]{IEEE_standard_for_definitions_of_terms_for_antennas_2013}.
Specifically, our framework enables the visualization of the convergence of the electromagnetic fields toward the far-field. 
While, in this paper, we propose a metric to quantify this convergence, we emphasize that alternative metrics may be more suitable depending on the target application.
Furthermore, we deliberately refrain from specifying a strict threshold at which the far field begins, since the IEEE standard does not define such a threshold either and the transition between the near-field and far-field regions is inherently of gradual nature. 
We have utilized our framework to experimentally evaluate a selection of existing single-letter threshold near–far-field boundaries.
These boundaries have been tested in various single- and multi-antenna wireless systems, using analytical models as well as full-wave electromagnetic simulations.
The results of our experiments clearly demonstrate that all of the considered single-letter distance thresholds are insufficient to accurately predict the transition region between the NF and FF regions.
Since the tested single-letter thresholds were not derived from first principles, but often from simplified geometric models, this finding highlights the importance of basing concepts related to antenna systems---such as near–far-field boundaries---on Maxwell’s equations. 
Additionally, throughout the paper, we have revisited and clarified several fundamental yet often misunderstood concepts related to NF and FF regions and antenna theory in general.
We have emphasized that the field regions are intrinsic properties of an antenna (\fref{rem:intrinsic_property}) and that, in multi-transceiver scenarios, each antenna system possesses its own NF and FF regions (\fref{rem:multiple_tranceivers}).
Furthermore, we have demonstrated that the common heuristic of identifying field regions based on plane- or spherical-wave characteristics lacks physical justification and can lead to incorrect conclusions (\fref{rem:misconseption_plane_spherical_wave}).
Likewise, we have highlighted that, in general, higher frequencies do not imply larger NF regions (\fref{rem:frequency_size}); 
that no antenna can behave as an isotropic radiator (\fref{rem:isotropic_radiator}); 
that the largest antenna dimension must not be confused with its physical aperture (\fref{rem:Dp_not_Ds}); 
that the Rayleigh distance is well-defined only for aperture antennas (\fref{rem:rayleigh_defined_for_aperture_antennas}); 
and that there is no physical justification for interpreting the quasi-Rayleigh distance---which is typically referred to as Rayleigh distance---as the boundary between the NF and FF regions of a general antenna (\fref{rem:do_not_use_pseudo_rayleigh_for_definition}).
Finally, we have clarified that the ``Björnson distance,'' introduced by Björnson \emph{et al.}, should not be interpreted as a near–far-field boundary (\fref{rem:bjoernson_distance}). 
Collectively, these insights contribute to a more physically consistent understanding of NF and FF concepts in modern wireless communication and sensing systems.

For reproducibility, the complete MATLAB code and Ansys HFSS project files we have used in our experiments are available at \url{https://github.com/IIP-Group/On_Near_Far_Field_Boundaries}.
%

\balance
\bstctlcite{IEEEexample:BSTcontrol} 
\bibliographystyle{IEEEtran}
\bibliography{bib/publishers,bib/IEEE_abbr,bib/journals_proceedings_ect,bib/library}
\balance

\end{document}